\documentclass[twocolumn]{aastex61} 
\pdfoutput=1 
\usepackage{hyperref} 
\usepackage[figure,figure*]{hypcap}
\hypersetup{
  colorlinks=true, citecolor=blue, linkcolor=blue, filecolor=magenta,
  urlcolor=blue }

\newcommand{\be}[1]{\begin{equation} \label{eq:#1}}
\newcommand{\ee}{\end{equation}}
\newcommand{\ba}[1]{\begin{eqnarray} \label{eq:#1}}
\newcommand{\ea}{\end{eqnarray}}

\newcommand{\CaII}{\ion{Ca}{2}}

\newcommand{\solrad}{\ifmmode{R}_{\rm S}\else${R}_{\rm S}$\fi}
\newcommand{\solmas}{\ifmmode{M}_{\rm S}\else${M}_{\rm S}$\fi}

\newcommand{\ctn}{\ifmmode\kappa\else$\kappa$\fi}

\newcommand{\flxu}{$\,$ergs$\,$cm$^{-2}\,$s$^{-1}$}

\newcommand{\fluxu}{\ifmmode{\rm erg~cm^{-2}~s^{-1}}\else 
  erg~cm$^{-2}$~s$^{-1}$\fi}
\newcommand{\velu}{$\,$km$\,$s$^{-1}$}

\newcommand{\wave}{\ifmmode{\lambda} \else$\lambda$\fi}

\newcommand\lta { \mathrel {\hbox to 0pt {\lower 3.7pt \hbox{$\sim$}
      \hss} \raise 1.7pt \hbox{$<$}}}
\newcommand\gta { \mathrel {\hbox to 0pt {\lower 3.7pt \hbox{$\sim$}
      \hss} \raise 1.7pt \hbox{$>$}}}

\catcode`_=\active
\newcommand_[1]{\ensuremath{\sb{\mathrm{#1}}}}

\newcommand{\revise}[1]{#1}

\newcommand{\rawdata}{ \protect\begin{deluxetable*}{lllllllll}
\vskip 24pt \tablecaption{Fundamental data for older main sequence stars  \label{tab:rawdata} }
\tablehead{Star &Sp& Age & $M$ &$R$ &$L^\dag$ & $P_{rot}$ & R.V.  &
  $\phi^\prime$\\ 
&type& (Gyr) & ($M_\odot$) & ($R_\odot$) & ($L_\odot$)
& (days) & (\velu{})
  &(mas)} 
\startdata 
18 Sco &G2~V &$3.7\pm0.5^a$&$1.03\pm0.01^a$& 1.01$^b$ & 1.06&22.7$^c$ &+11.6 & $0.6797\pm0.0062$\\
 Sun &G2~V& $4.6 \pm0.04^d$&1 &1 & 1 &25.4 & 0 & \ldots \\
 $\alpha$ Cen A & G2~V&$5.4^{+1.2}_{-0.2}$$^e$&$1.11\pm0.01^e$& $1.227^f$ &1.53 &22--29$^g$ &$-$23.45& $8.662 \pm 0.02$ \\
16 Cyg A & G1.5 Vb & $7.0\pm0.3^h$ &$1.08\pm0.02^h$ & $1.194^i$ &1.58& 23.8$^j$ & $-$27.5& $0.539\pm0.007$ \\
16 Cyg B &G3 V & $7.0\pm0.3^h$ &$1.04 \pm0.02^h$& $1.108^i$ &1.24 &23.2$^j$ & $-$28.1 & $0.481 \pm 0.006$ \\
$\tau$ Ceti &G8 V & $7.6^{+0.9}_{-1.5}$$^k$ &$0.78\pm0.01^l $& $0.793^l$ & 0.50& 34$^m$ & $-16.4\pm0.9$ &$2.03 \pm 0.05$ \\
\enddata
\tablecomments{Spectral types and radial velocities (R.V.)  are from
  the SIMBAD database.  $^\dag$ Luminosities are from \citet{Valenti+Fischer2005}. 
References:
  $^a$\cite{Li+others2012}, $^b$\citet{Bazot2011}, $^c$\cite{Petit2008},
  $^d$\cite{Houdek+Gough2011}, $^e$\cite{Bazot+others2016}, $^f$\cite{Kervella2003},
  $^g$\cite{Mamajek2008}, $^h$\cite{Metcalfe+others2015}, 
$^i$Adopted from seismic data of \cite{Buldgen+others2016}, supported by 
interferometric work of \cite{White+others2013} who found 
radii of $1.22\pm0.02$ and $1.12\pm0.02 R_\odot$ for 16 Cyg A and B respectively.  
$^j$\cite{Davies2015},
  $^k$\cite{Pagano+2015}, $^l$\cite{Tang+Gai2011}, $^m$\cite{Baliunas1996}.  The
  angular diameters $\phi^\prime$ were derived from {\em Hipparcos}
  distances and listed radii. }
\end{deluxetable*}
}

\newcommand{\tabobs}{ \protect\begin{deluxetable}{llllll}
\vskip 24pt \tablecaption{Stellar data from HST/COS \label{tab:obs} }
\tablehead{ Target & COS Grating$^\dag$ & Start Time & Exposure &
  $\lambda$ range \\
 & & UTC & sec & nm} \startdata 18 Sco & G130M &
2011-02-04 21:46 & 2700.54 & 115-145\\ 16 Cyg A & G130M & 2015-10-23
01:58 & 11677.8 & 115-145\\ & G160M & 2015-10-23 08:21 & 5882.6&
140.5-177.5\\ 16 Cyg B & G130M & 2016-02-03 01:10 & 11674.5 &
115-145\\ & G160M & 2016-02-03 07:31 & 5882.4 & 140.5-177.5\\ \enddata
\tablecomments{ 16 Cyg data are from program 13861.  18 Sco data are
  from program 12303.  $^\dag$The observation IDs are LCN501010 (for
  G130M) and LCN501020 (G160M) for the 16 Cyg stars, and LBIZ01020 for
  18 Sco.}
\end{deluxetable}
}

\newcommand{\figsts}{
\begin{figure}
\centering \includegraphics[width=0.96\linewidth]{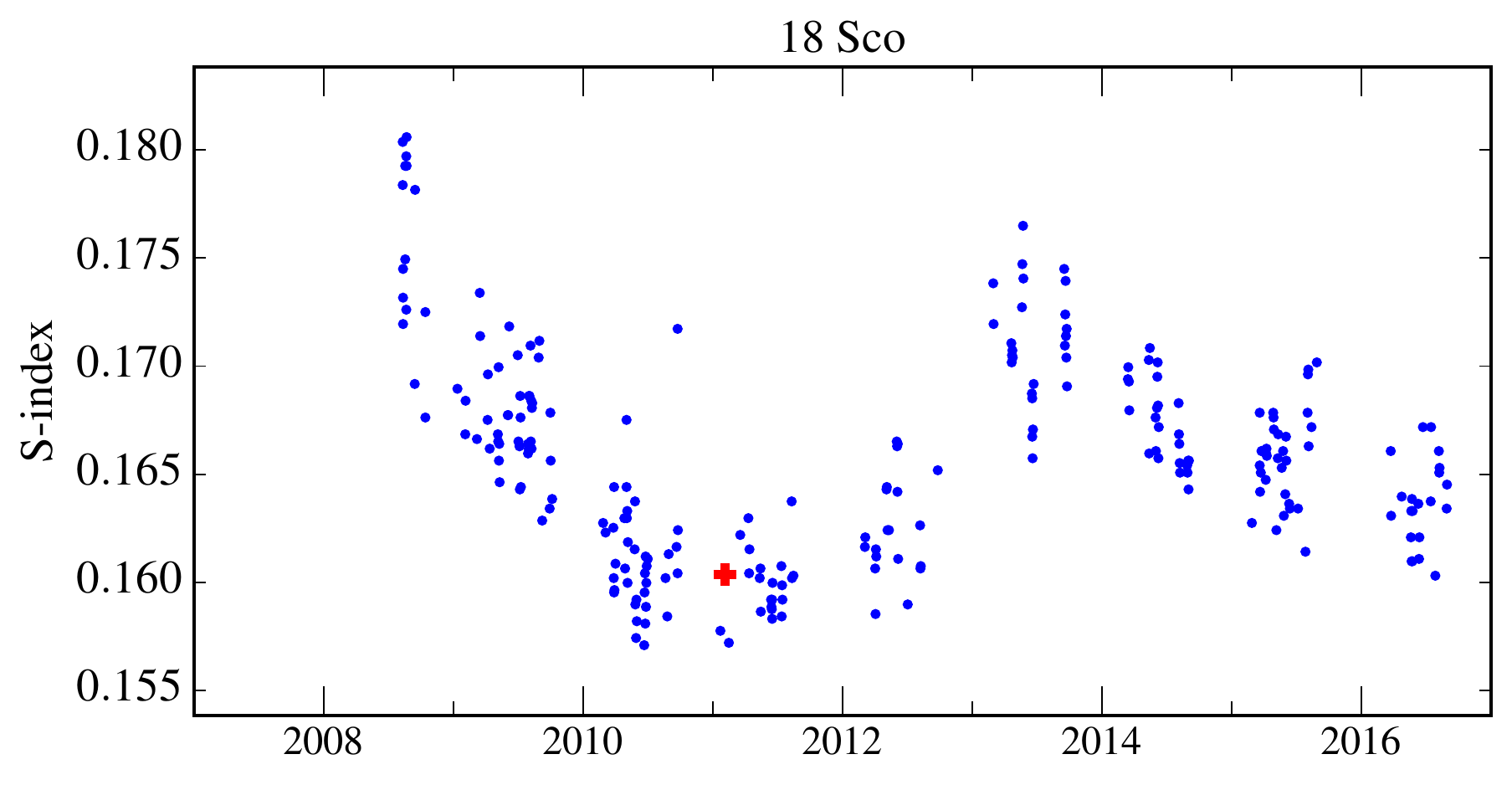}
\includegraphics[width=0.96\linewidth]{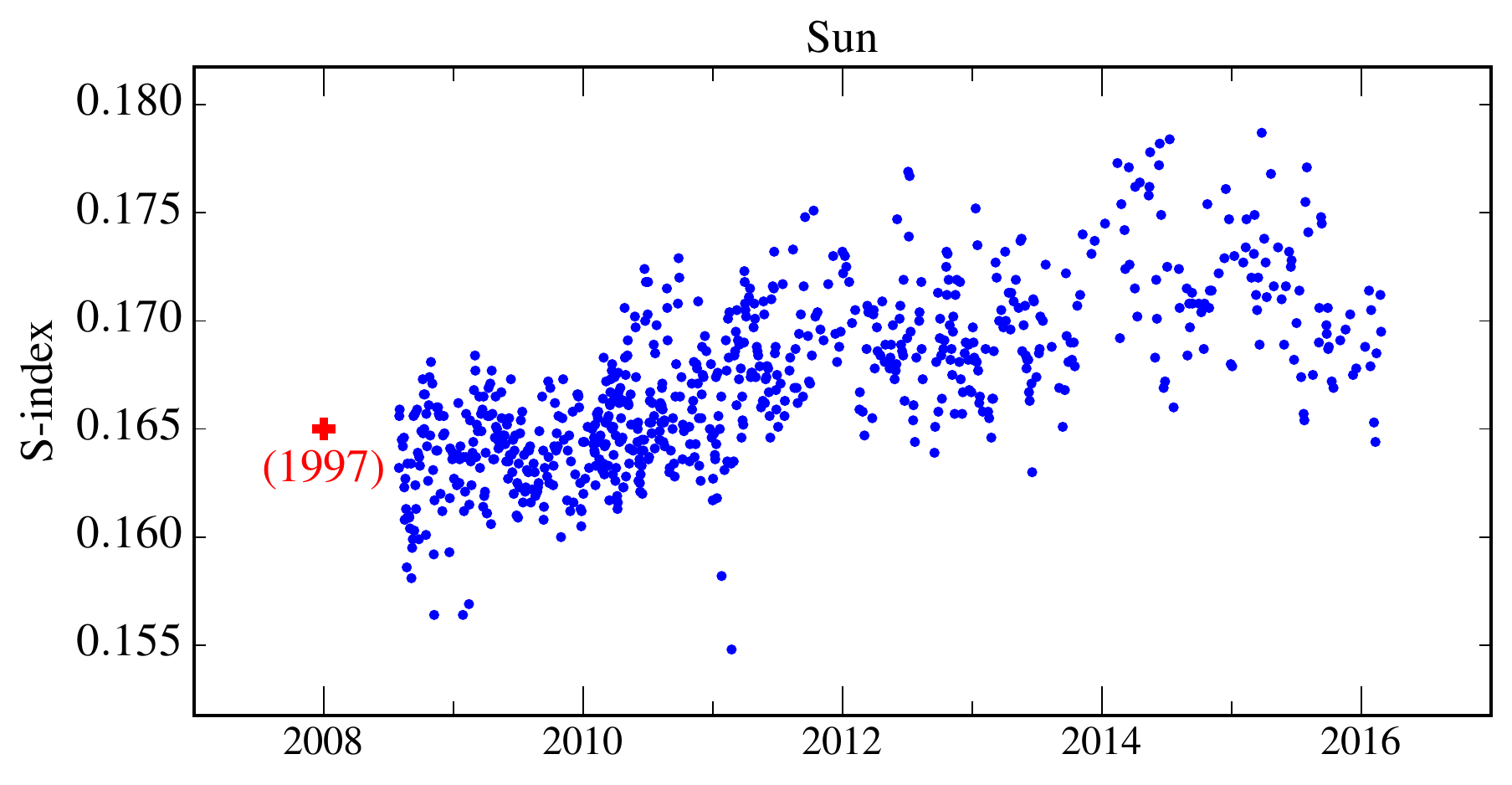}
\includegraphics[width=0.96\linewidth]{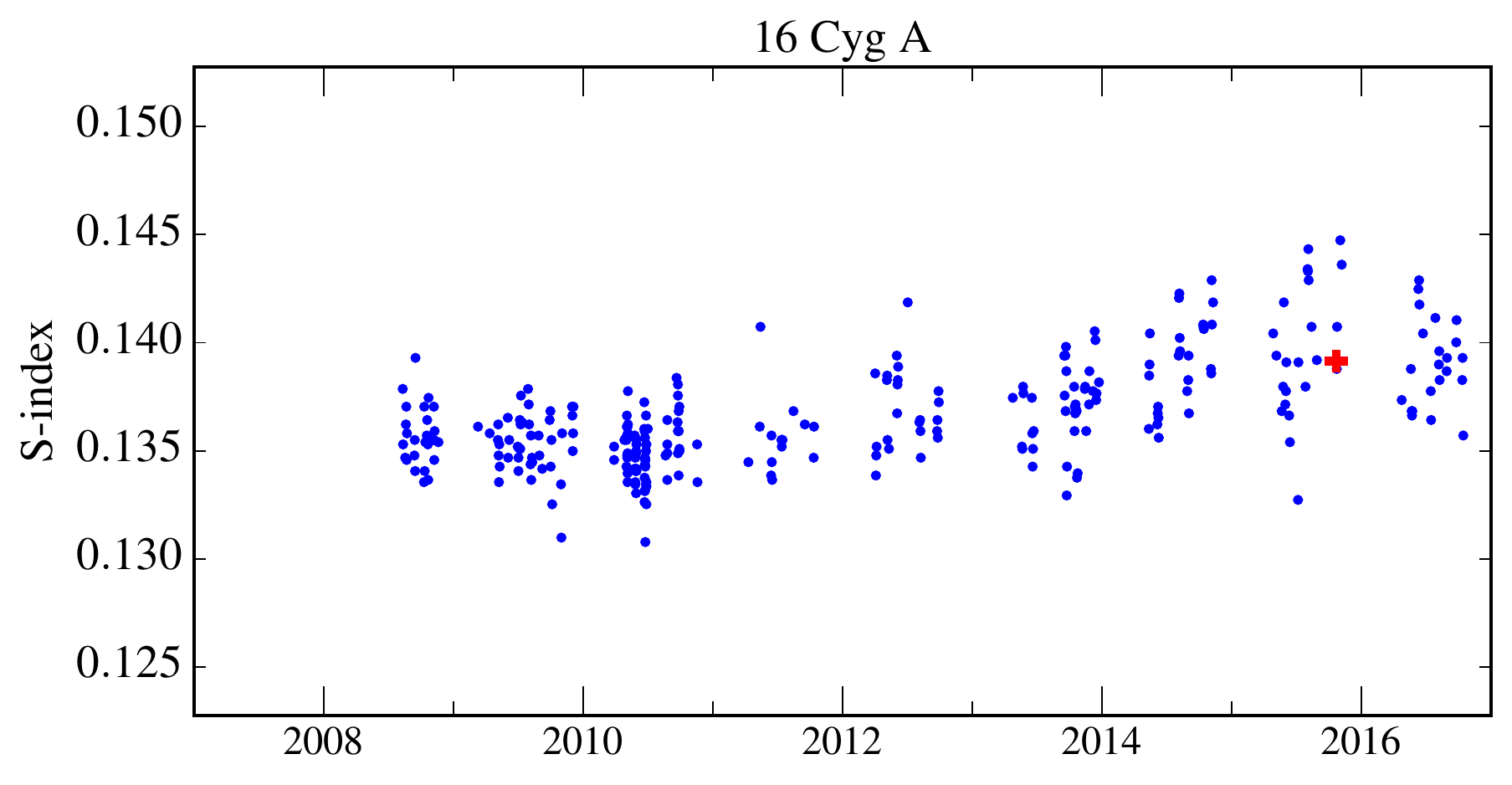}
\includegraphics[width=0.96\linewidth]{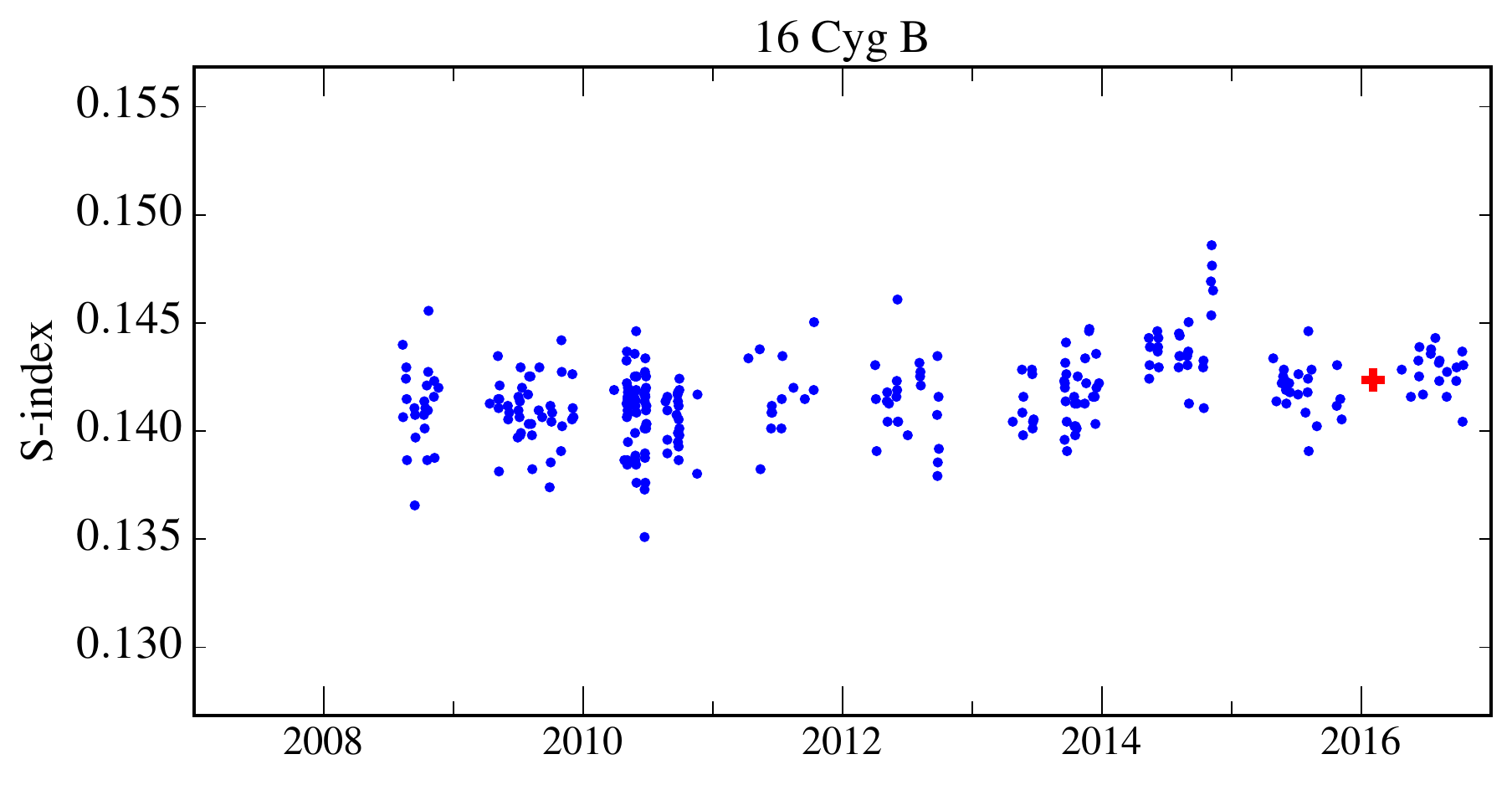}
\includegraphics[width=0.96\linewidth]{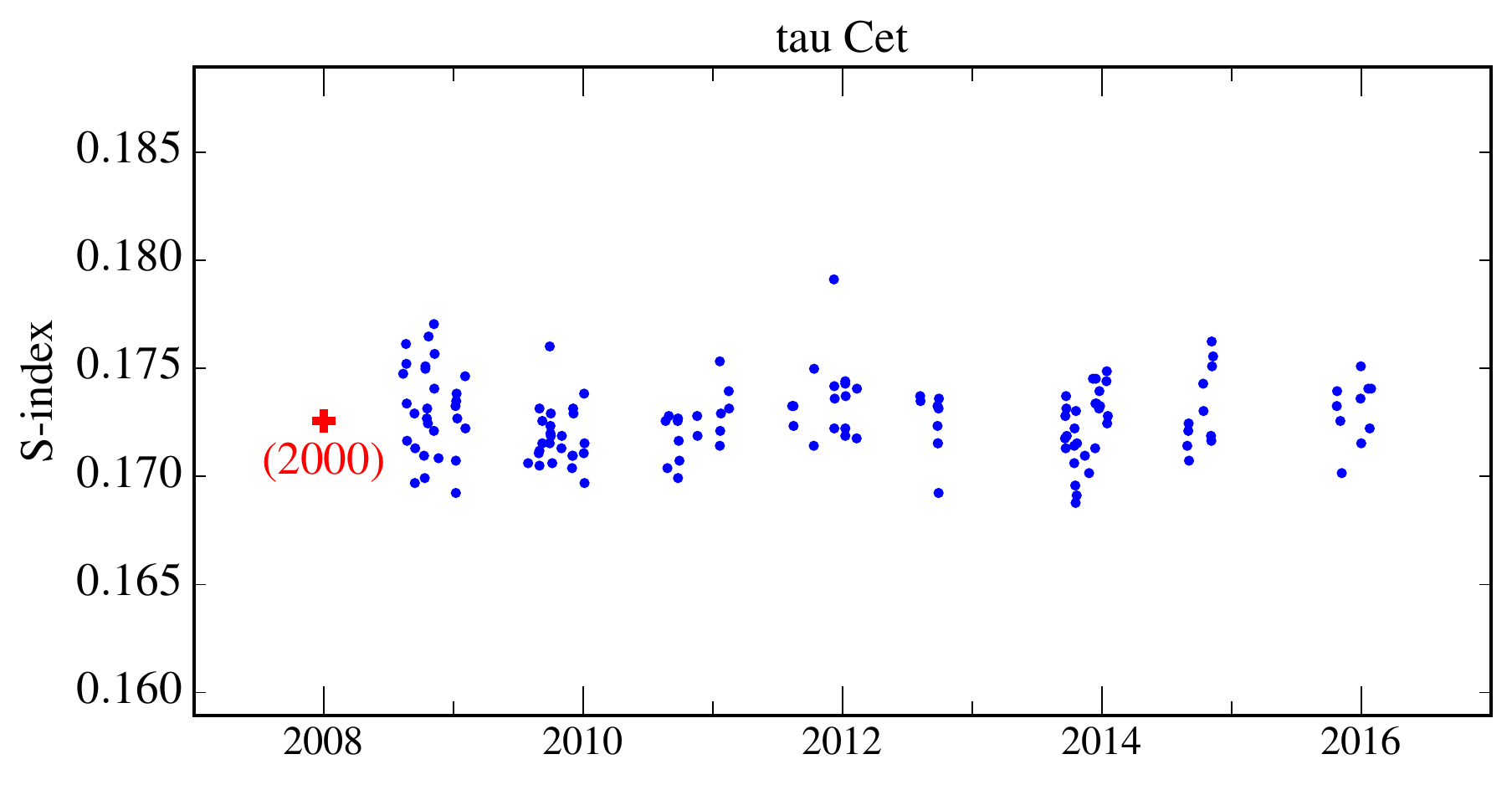}\caption{Solar Stellar Spectrograph $S$-index data for 18 Sco, the
  Sun, 16 Cyg A and B \revise{and $\tau$ Cet are shown, from  \cite{Egeland+others2017}.
Each plot spans a range of 0.035 in S, so that amplitudes can be directly compared.
 Red crosses mark the time of the COS
 UV stellar observations and the 2-year $S$-index median centered at that time. The solar UV  data were obtained
close to solar minimum on April 20, 1997.  STIS data for $\tau$ Cet were obtained on August 1 2000.  These two data 
points are arbitrarily plotted at year 2008.}}
\label{fig:Sts}
\end{figure}
}

\newcommand{\figSindex}{
\begin{figure}
\centering \includegraphics[width=\linewidth]{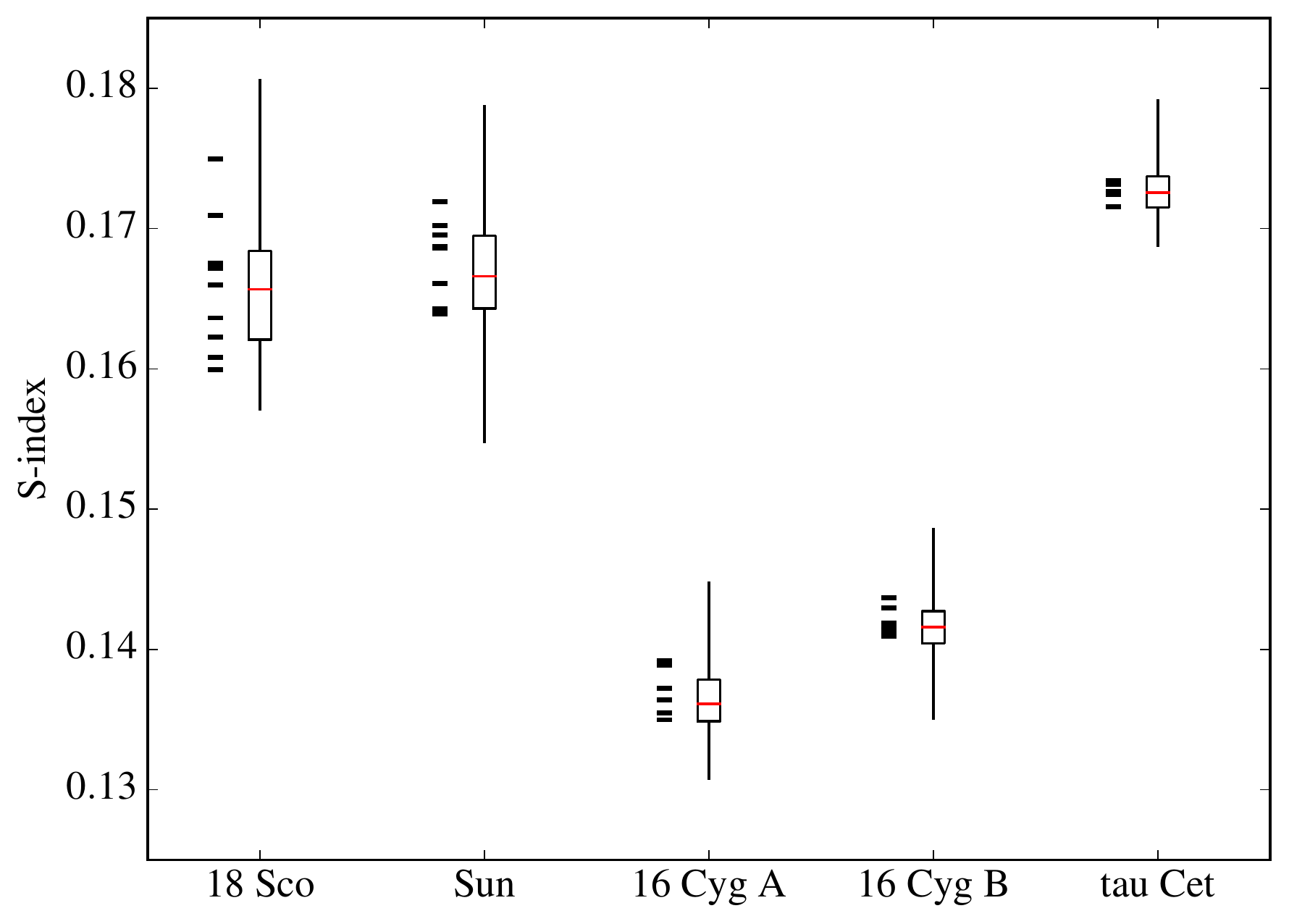}
\caption{Solar Stellar Spectrograph $S$-index variability
  from 2008 to 2016. The stars are listed in order of increasing age.
Black dashes show
  seasonally binned medians, or 1-year binned medians for the Sun.
  The box extends from the lower to upper quartile values of the data,
  with a line at the median. The
  whiskers extend from the box to show the total range of the data. \revise{The boxes span a range in $S$-index equivalent to 1.47$\sigma$, where $\sigma$ is the width parameter in a normal distribution}.}
\label{fig:Sindex}
\end{figure}
}

\newcommand{\figall}{
\begin{figure*}
\includegraphics[width=\linewidth]{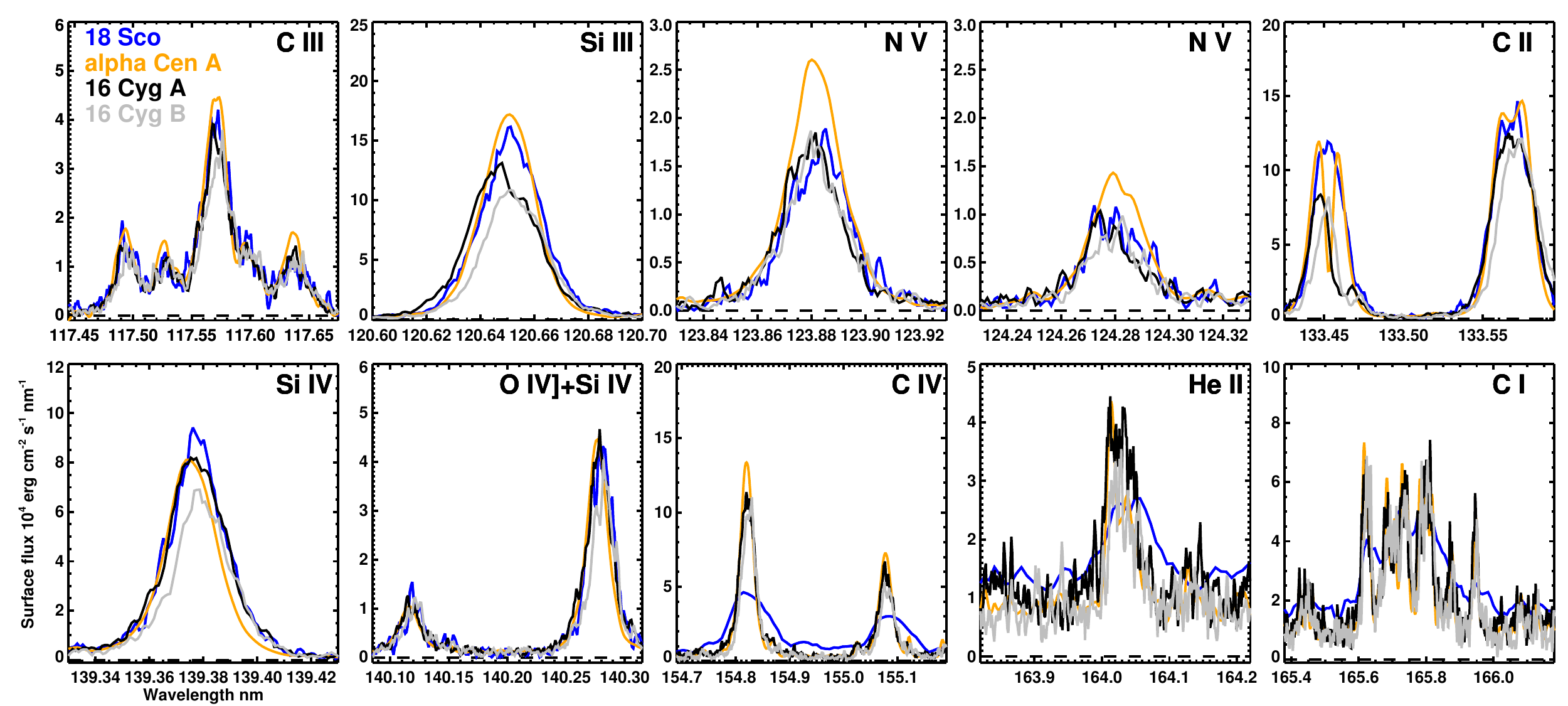}
\caption{Surface flux densities ${\cal F}_{UV}$ of prominent UV lines in the spectra of the
  stars observed are shown (not including disk-center data for the Sun
  or $\tau$ Ceti).  The measured flux densities are normalized to unit
  area of the stellar surface in \fluxu{}~nm$^{-1}$, and the wavelengths have been
  set to the photospheric reference frame using the relative
  velocities given in Table~\ref{tab:rawdata}. 
  \revise{The uncertainties in flux density from these HST UV data are $\pm 5\%$.}
  Note that data for
  $\alpha$ Cen A are from the STIS instrument.  Interstellar
  absorption is present in the 133.45 nm line of \ion{C}{2}.  In the
  panel for \ion{O}{5}], the ``continuum'' is a mix of geo-coronal and
stellar L$\alpha$ emission.}
\label{fig:all}
\end{figure*}
}

\newcommand{\figsun}{
\begin{figure*}
\includegraphics[width=\linewidth]{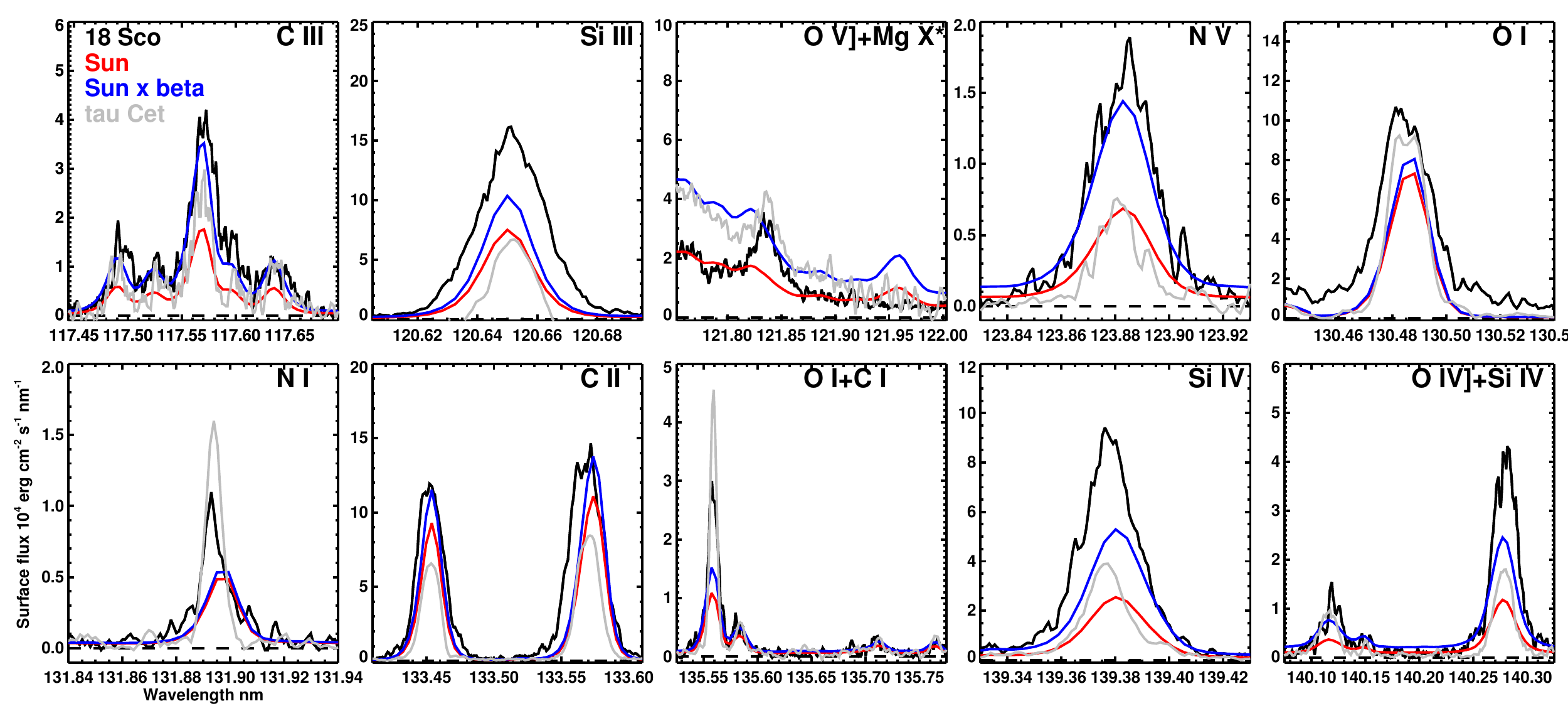}
\caption{Flux densities ${\cal F}_\lambda$ of 18 Sco are shown with
  disk-center data for the quiet Sun.  The disk center intensity has
  been multiplied by $\pi$ to produce a solar flux spectrum that
  assumes zero limb-darkening or brightening (red lines).  The blue lines 
have been multiplied by the additional limb-brightening 
factors $\beta_i$ discussed in the text.
 }
\label{fig:sun}
\end{figure*}
}

\newcommand{\figyoung}{
\begin{figure}
\includegraphics[width=\linewidth]{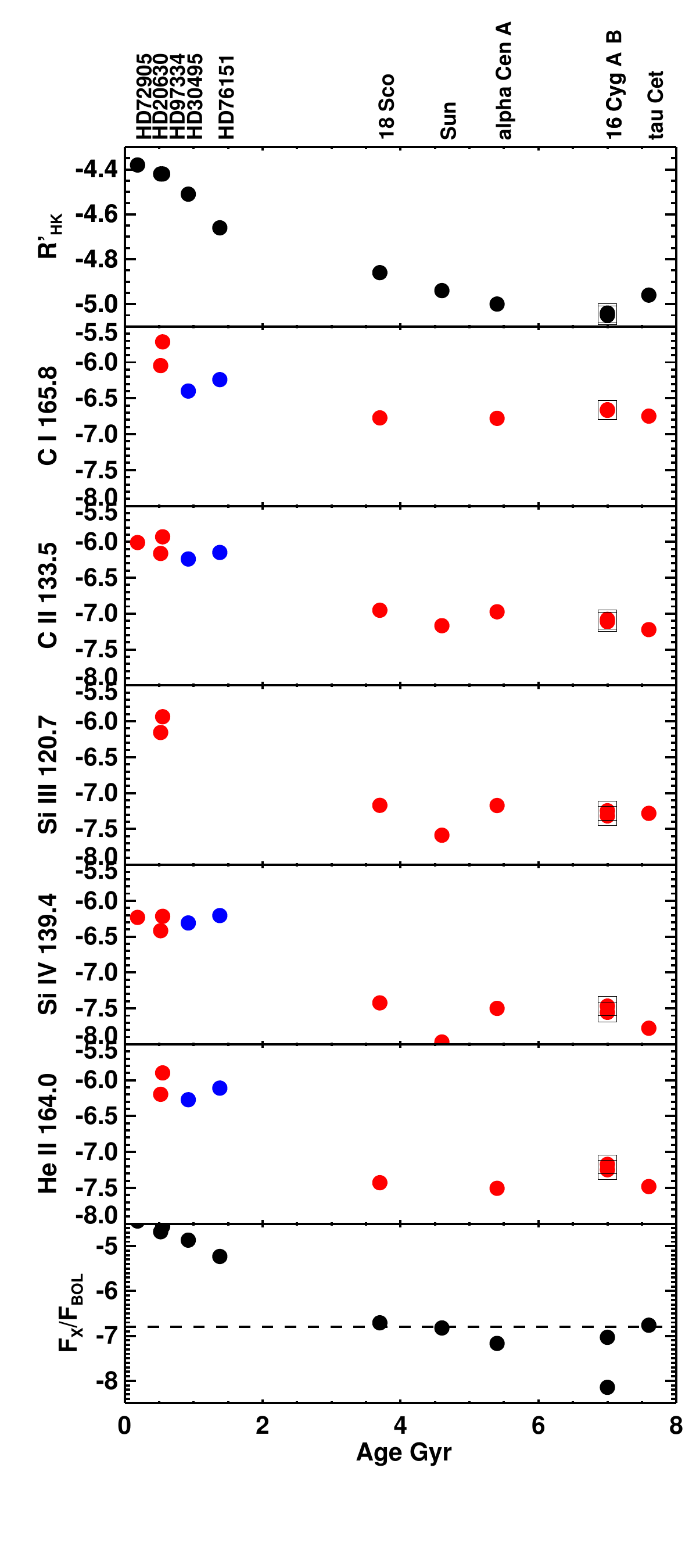}
\caption{Logarithmic flux densities for stars from 0.2 to 7.6 Gyr ages, normalized
  to the stellar bolometric flux.  Blue dots show poorer quality
  spectra from the IUE satellite, red dots data from the HST. The
  squares highlight 16 Cyg A and B.  The UV solar data plotted show the 
uncorrected disk center intensities, which are factors $\beta_i$ below the stellar data owing to the absence of a solar flux spectrum (see text).
X-ray data in the lowest panel are from sources in the text.  The lowest point is for 16 Cyg B which has $\log_{10} L_X = 25.5$.
\revise{Relative uncertainties are dominated by those in the absolute flux calibrations for the UV lines, at most 14\%{} 
(IUE vs. IUE), 7\%{} (HST vs. HST) and 11\% (IUE vs. HST}). 14\% corresponds to 
0.057 in the logarithm, 2-3$\times$ smaller than the diameters of the symbols. 
}
\label{fig:young}
\end{figure}
}

\newcommand{\fignl}{
\begin{figure}
\includegraphics[width=\linewidth]{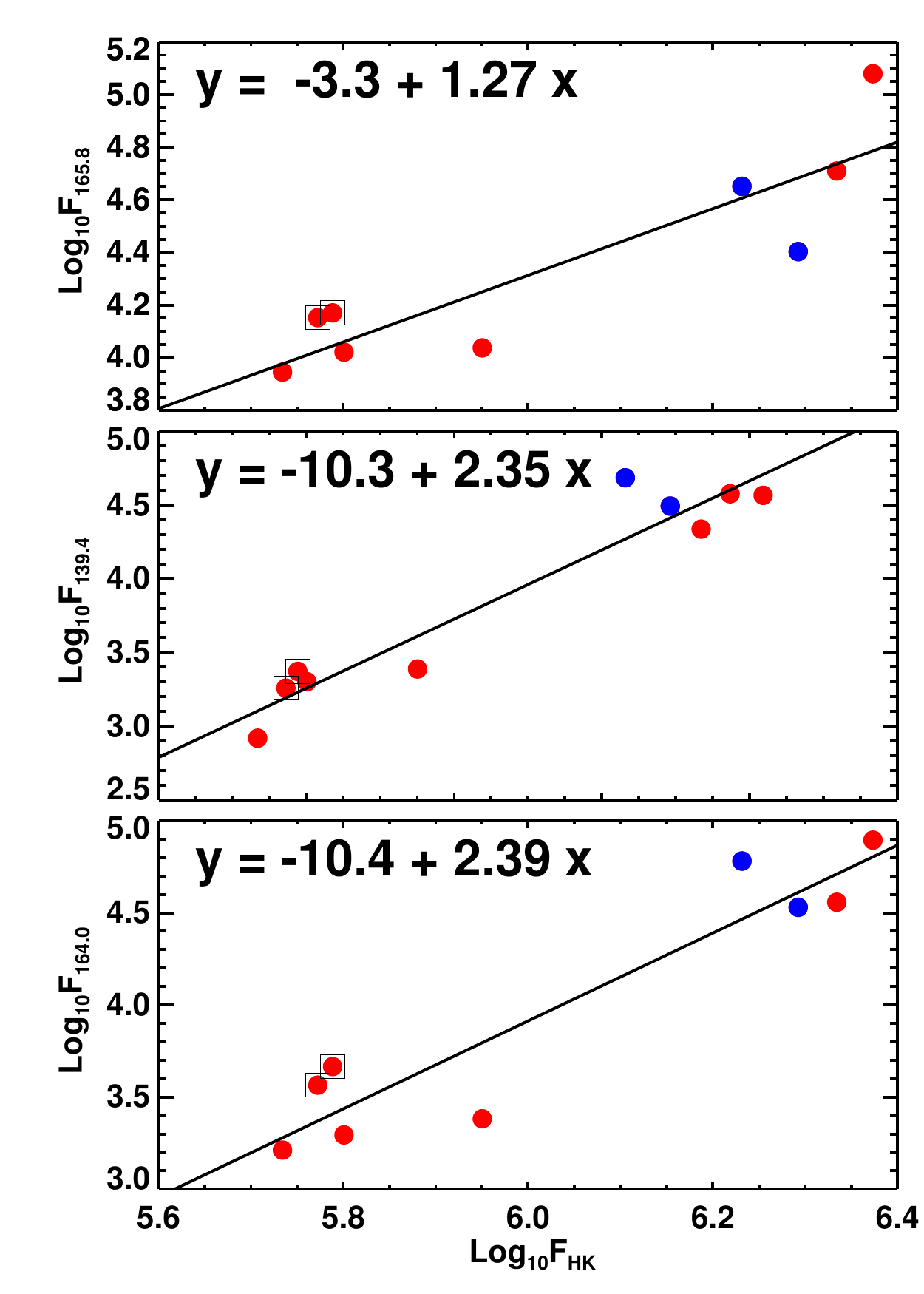}
\caption{
Scatter plots of \revise{surface flux densities ${\cal F}$} for stars from 0.2 to 7.6 Gyr
  ages.  The upper panel
  shows the \ion{C}{1} 165.7 nm multiplet plotted against
  ${\cal F}_{HK}$, the lower the 139.3 nm line of \ion{Si}{4} and the
164.0 nm (B$\alpha$) multiplet of
  \ion{He}{2}. The latter is somewhat sensitive to coronal irradiation.  As
  for Figure \protect\ref{fig:young}, the red dots show data from HST,
  the blue dots show (noisier) data from the IUE satellite.  \revise{As in Figure~\ref{fig:young}, the error bars are smaller than the symbols. }
\label{fig:nl}}
\end{figure}
}

\newcommand{\figdeconv}{
\begin{figure*}
\includegraphics[width=\linewidth]{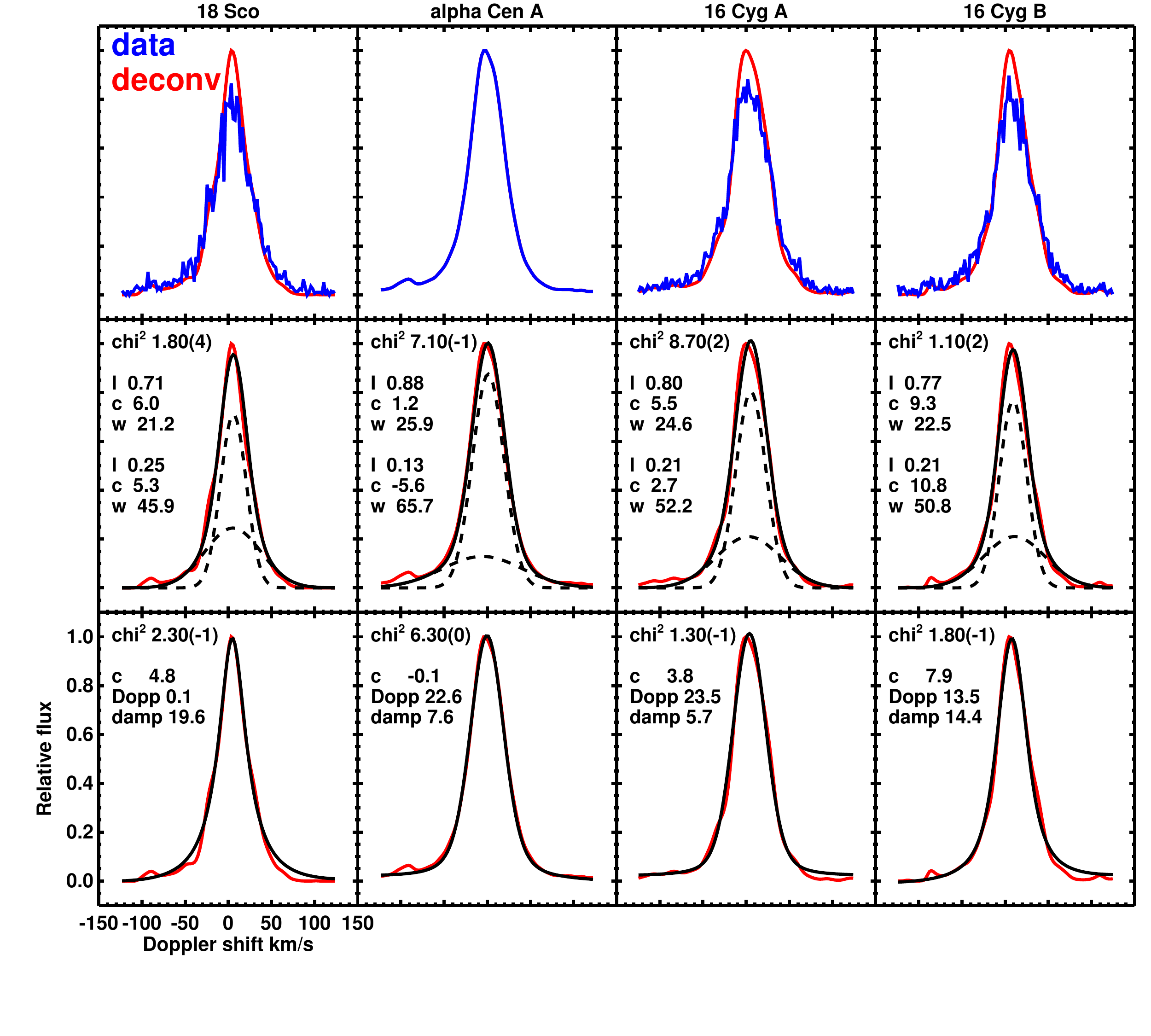}
\caption{Raw and deconvolved profiles are shown in the upper panels, and fits to the deconvolved profiles are shown in the lower panels, for the 
139.3755 nm line of \ion{Si}{4} The data for $\alpha$ Cen A are from the STIS instrument for which deconvolution is not necessary. \revise{``chi$^2$'' is the reduced $\chi^2$, with notation 1.80(4) $\equiv 1.80\times10^4$.  Two sets of values for the peak relative intensity $I$ and line center $c$ and width  $w$ are listed for the two Gaussian components in km/s, and Doppler and damping parameters are listed for the single Voigt component.  }}
\label{fig:deconv}
\end{figure*}
}

\newcommand{\figcl}{
\begin{figure}[ht]
\begin{center}
\includegraphics[width=1\linewidth]{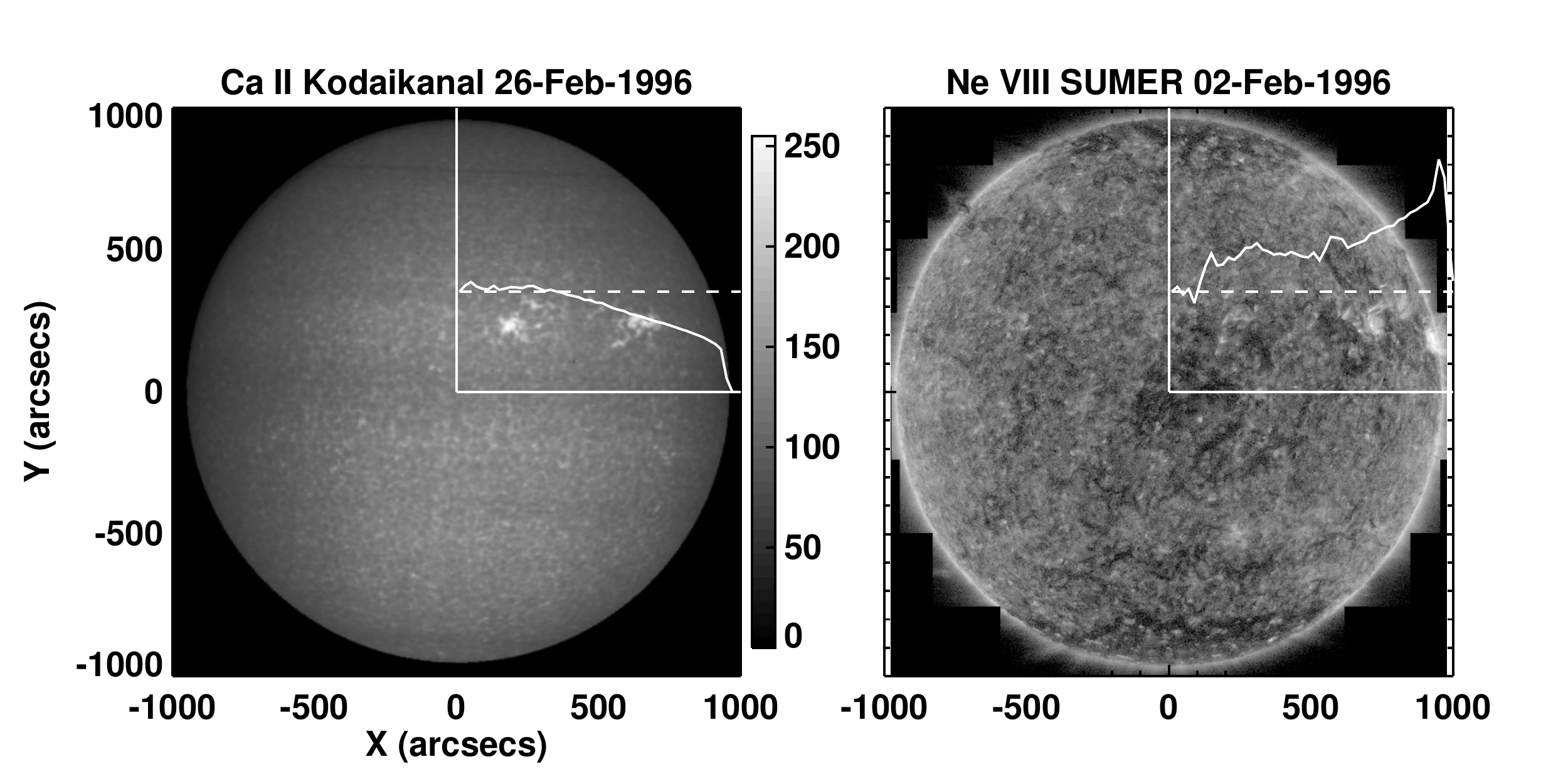}
\end{center}
\caption{Intensity images are shown for the core and inner wings of
  the \ion{Ca}{2} $K$ line (from the Koidaikanal observatory archive)
  and in the \ion{Ne}{8} coronal line from the SUMER instrument.  The
  two images are from different days, during a very low activity
  period in February 1996.  The line figures show the average
  center-to-limb behavior computed from the sector of quiet Sun
  between 7:30 and 10:30 on a clock-face.  The dashed line shows no
  center-to-limb variation.  The two curves are qualitatively
  different and the coronal line extends above the visible light limb.
}
\label{fig:cl}
\end{figure}
}

\newcommand{\figpassage}{
\begin{figure}[ht]
\begin{center}
\includegraphics[width=0.9\linewidth]{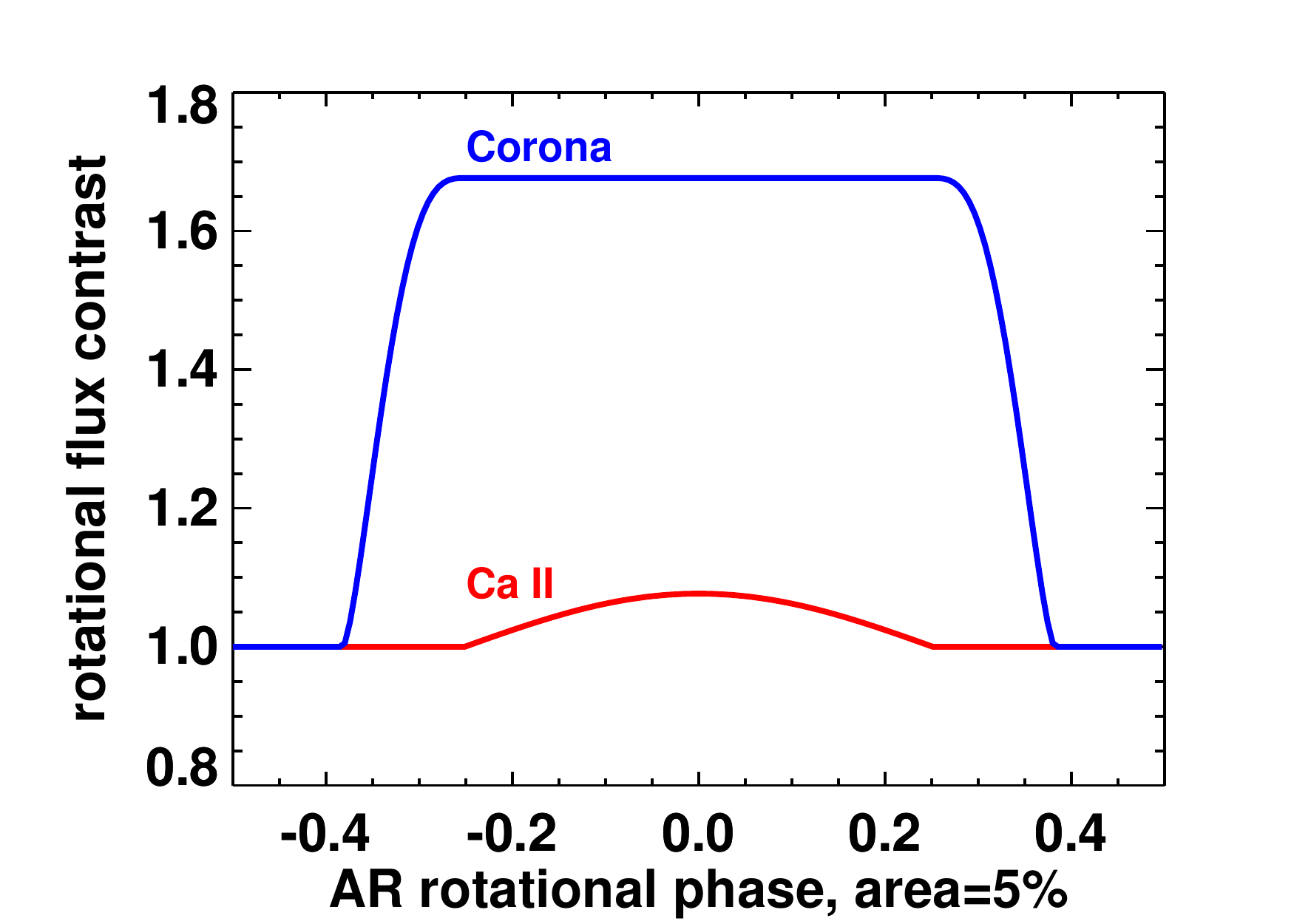}
\end{center}
\caption{Computed rotational modulation of the total (disk-integrated)
  \ion{Ca}{2} S index and optically thin coronal emission are shown as
  a function of rotational phase of an active region as it passes
  across a stellar disk.  The background atmosphere is limb-darkened
  at \ion{Ca}{2} wavelengths (using the parameters of
  \citealt{Allen1973}) and limb-brightened, and extended across the
  limb by a pressure scale height (0.06 of the stellar radius) in the
  coronal bandpass.  The active region covers 5\% of the disk area,
  and is $2\times$ (\ion{Ca}{2}) and $100\times$ (corona) brighter
  than the background disk-center intensity respectively. The disk-integrated \ion{Ca}{2}
  has a 10\% increase when the active region crosses central meridian.
  In spite of the small area and modest increase in brightness, the
  modulated signal in \ion{Ca}{2} is 10\%. The disk-integrated coronal modulation is
  just 70\% in spite of the fact that the the AR is 100$\times$
  brighter than the surrounding corona, because it adds to a strongly
  limb-brightened disk and emission from above the limb.  Coronal AR
  emission is isotropic, once it transits the disk the flux is
  constant until it goes into eclipse.  }
\label{fig:passage}
\end{figure}
}

\shorttitle{Solar magnetic future} \shortauthors{Judge et al.}

\begin{document}

\title{The magnetic future of the Sun}

\author{Philip G. Judge$^1$, {Ricky Egeland}}
\affil{{High Altitude Observatory, National Center for Atmospheric Research\footnote{The National Center for Atmospheric Research is sponsored by the National Science Foundation}, P.O.~Box 3000, Boulder CO~80307-3000, USA}}
\author{Travis S. Metcalfe}
\affil{Space Science Institute, 4750 Walnut Street, Suite 205, Boulder CO 80301, USA}
\author{Edward Guinan,$^3$ Scott Engel}
\affil{Astronomy and Astrophysics Department, Villanova University, 800 Lancaster Ave, Villanova PA 19085, USA}

\begin{abstract}
\revise{We analyze space- and ground-based data for the old ($7.0\pm0.3$~Gyr) solar analogs 16 Cyg A and B.  The stars were
observed with the Cosmic
Origins UV Spectrographs on the Hubble Space Telescope (HST) on 23 October
2015 and 3 February 2016 respectively,
and with the Chandra X-ray Observatory on 7 February 2016. Time-series data in \ion{Ca}{2} data are used to place the UV data in context.  The UV spectra of
18 Sco (3.7$\pm0.5$ Gyr), the Sun (4.6$\pm0.04$
Gyr) and $\alpha$ Cen A ($5.4_{-0.2}^{+1.2}$ Gyr), appear remarkably similar, pointing to a convergence of 
magnetic heating rates for G2 main-sequence stars older
than $\approx 2-4$ Gyr. 
But the B component's X-ray
(0.3-2.5 keV)
flux lies 20$\times$ below 
a well-known minimum level reported by Schmitt. As  reported for $\alpha$~Cen~A,  the coronal temperature probably lies below that detectable in soft X-rays.  No solar UV flux spectra of comparable resolution to stellar data exist, but they are badly needed for comparison with stellar data.  
Center-to-limb (C-L) variations are  
re-evaluated for lines such as \ion{Ca}{2} through to X-rays, with important consequences 
for observing activity cycles in such features.  We also call into question work that has mixed solar intensity-intensity statistics with flux-flux relations of stars.  }
\end{abstract}


\section{Introduction}
 
Ultraviolet emission from the Sun was originally inferred by the
existence of the ionosphere
\citep[e.g.][]{Chapman1931,Woolley+Allen1948}. Solar UV radiation arises primarily because
mechanical energy from the solar interior is deposited into the Sun's tenuous atmosphere, in structures from which UV photons can escape.
Solar observations from the space era reveal clear evidence that 
the solar magnetic field is the agent responsible for heating
most of the plasma capable of emitting UV photons that can escape and
reach the Earth \citep[as reviewed by][]{Eddy2009}.

Solar magnetism exhibits remarkable properties.  The Sun acts as a
magnetic machine. A small fraction of the energy from nuclear fusion
is partitioned into magnetic energy, which is manifested in two
seemingly universal phenomena entirely unanticipated from first
principles \citep[e.g.][]{Judge2017}.  Firstly, it regenerates its
entire magnetic field every 22 years, and secondly, it converts a
small but (for life on Earth) significant amount of energy into high
energy photons and particles flooding the solar system.

\figsts

We have yet to find a Sun-like star that does not emit soft X-rays
\citep{Schmitt1997}.  The Sun's outer layers, sitting atop an
atmosphere with energies close to 0.4 eV (4500 K), appear to emit UV
radiation (10 eV), X-rays (100-5000 eV), even occasionally
$\gamma$-rays (MeV) during flares.  The varying magnetic field in
Sun-like stars produces near-UV emission in the cores of
Fraunhofer's $H$ and $K$ lines of \ion{Ca}{2}
\citep{Leighton1959,Linsky+Avrett1970}.  Such changes, readily observable from the
ground, have been cataloged for more than 5 decades
\citep{Wilson1968,Wilson1976,Baliunas+others1995,Hall+Lockwood2004,Hall+others2009,Egeland+others2017}.
Examples of this behavior for 18 Sco, 16 Cyg A and B, are shown in Figure~\ref{fig:Sts}.
The Sun-as-a-star shows modest variations in the $H$ and $K$
lines, somewhat less than the variations shown for 18 Sco 
\citep{Egeland+others2017}. All such variations are clearly
associated with surface magnetism
\citep{Skumanich+Smythe+Frazier1975}.  These and other observations
indicate that the Sun is not alone in showing its magnetic
fingerprints, decadal variations are common amongst late-type stars on
the main-sequence \citep{Baliunas+others1995}.  Here we study new UV
line profiles for two optically ``bright'' stars that are older
than the Sun, 16 Cyg A and B, \revise{obtained with the Cosmic Origins Spectrograph (COS) on the Hubble Space Telescope (HST)}.  These stars are among the weakest activity 
main-sequence sun-like stars found.  Earlier generations of UV spectrographs were
not sensitive enough to obtain useful data.  The FUSE spacecraft did
detect EUV emission lines (C III 977, O VI 1031) in the A component,
but the S/N ratios were insufficient to probe properties of line
profiles \citep{Guinan+others2003} that dynamically reflect magnetic
field activity in the atmospheres of stars and the Sun.

Our main purpose here is to study properties of the UV emission lines
(formed between $\approx 8000$ to $200,000$ K) with age, extending previous studies of bright young targets to relatively dim and old stars in the 16 Cygni system.
The emission features of doubly- and triply- charged ions in
Sun-like stars reflect magnetic heating mechanisms through their
mere presence \citep{Athay+White1978,Athay+White1979b}.  Hydrodynamic
processes alone also fail to account for even the most basic
properties of line profiles of the ``chomospheric'' \ion{C}{2} 133.5
nm multiplet \citep{Judge+Carlsson+Stein2003}.  X-ray studies show
that coronae are present on all Sun-like stars in a volume limited sample 
\citep{Schmitt1997}.    Coronae  must be
  powered by magnetic fields emerging from beneath the stellar
  surfaces, as must the broad line 
components observed in $\alpha$ Cen A and the Sun. Specifically, we study the \ion{Ca}{2} $S$-index time series, the UV line profiles from the Hubble Space telescope, including both core and broad components, as well as available X-ray data.  

The most
striking feature we find, is the remarkable similarity in the UV spectra of all
the G2 stars.  We show that this agreement is, at first inspection,
quite ``unreasonable'', given what we know of stellar flux-flux
relationships and the known variations of the \ion{Ca}{2} $S$-indices
for these stars.  A resolution of this difficulty is proposed.  We
speculate on implications for the future surface magnetism of
the Sun and other G-type stars for the latter portions of their main
sequence lifetimes.

\rawdata

\section{Analysis}

Table~\ref{tab:rawdata} lists basic data for the \revise{Sun and five} stars analyzed
here.  The age of $\tau$ Cet is near 7.6 Gyr, but its lower mass
of 0.78~M$_\odot$, separates it from the other Sun-like stars
discussed below.  \revise{We included the star because so few HST spectra of old GV stars exist, and it has been discussed as a proxy for the Sun in a grand minimum state \citep{Judge+others2004b}.}

\subsection{New observations}

COS observations were acquired under program 13861 (E. Guinan, P.I.).
The circumstances of these observations are given in
Table~\ref{tab:obs} and these epochs are marked in
Figure~\ref{fig:Sts}. These data were processed using Version 3.3 of
the IDL procedure {\tt
 COADD\_X1D}\footnote{http://casa.colorado.edu/home/$\sim$danforth/science/cos/}
using default parameters.  We used HST data for stars other than these
and 18 Sco, processed and described by \cite{Judge+others2004b}.
G130M Data for 18 Sco were downloaded from the MAST archive and
processed as for the 16 Cygni stars. No G160M or other moderate
resolution HST data exist for 18 Sco.  The 18 Sco COS spectra used
here are from program 12303 (E. Guinan, P.I., see Table
\ref{tab:obs}).  \revise{The absolute flux calibration of the HST UV spectrographs is known to better than 5\%{}
\citep{Pagano+2015}. For later reference, the flux calibration for low resolution (6\AA) spectra from the International Ultraviolet Explorer  is  $\le$ 10\%{} \citep{IUE2001}.}

\tabobs

\subsection{The HST data in the context of magnetic variations}

In Figure~\ref{fig:Sts} time-series of the \ion{Ca}{2} ``$S$-index'' for
the 16 Cyg stars and 18 Sco, are marked with the epochs of the
observations with COS.  The solar data are shown for the same period,
also showing the $S$-index when the partial-disk 
observations were acquired (April 1997, 
\citealt{Curdt+others2001}).  The solar
data are from a typical region, averaging 
 data from along SUMER's $1\times300"$ slit. 
This region samples just 
0.01\% of the solar area, extending over a region covering $\cos \vartheta \approx 1$
to $0.988$,
where $\vartheta$ is the angle between local  vertical on the Sun and line-of-sight. 

We emphasize that the solar data shown are 
\emph{intensity} data, i.e. from 
a small, resolved part of the solar disk. The stellar data are \emph{flux} spectra.  \emph{A high resolution UV flux
spectrum of the Sun does not exist.} These 
differences will be important below. 

The \ion{Ca}{2} data in Figure~\ref{fig:Sts} were obtained from the Lowell Observatory
Solar-Stellar Spectrograph from 2008 to 2016, following an upgrade of
the instrument CCD detector.  The instrument and data reduction are
described by \cite{Hall+others2007a} with updates in
\cite{Egeland+others2017}.  The data were calibrated to the Mount Wilson Observatory instrumental scale using near-coincident measurements, as described in \cite{Egeland2017}.  The ``$S$-index'' measures the brightness of
the chromospheric cores of the \ion{Ca}{2} resonance lines relative to
the neighboring line wings \citep[e.g.][]{Wilson1968,
  Vaughan+Preston+Wilson1978}. \ion{Ca}{2} core emission is a
well-known ``proxy'' for the amount of magnetic activity in late-type
stars \citep{Leighton1959, Skumanich+Smythe+Frazier1975}. From
Figure~\ref{fig:Sts}, it is clear that 18 Sco was observed with COS close to a
minimum in $S$-index. The 16 Cyg stars have no clear cycle such as that
seen in 18 Sco with a period close to 7 years, or the Sun with a
period close to 11 years.  The solar measurements are also from a
period close to a minimum $S$-index, but these measurements are from
quiet regions close to disk-center only.

\figSindex

Figure \ref{fig:Sindex} summarizes long-term variability in the $S$-index shown in Figure \ref{fig:Sts}.  The range of $S$-index variability decreases with age, as can be seen
by the spread in the seasonal mean values as well as the interquartile
range.  The Sun's larger total range may be enhanced by the
significantly higher sampling cadence.  Using the more robust seasonal
medians and interquartile range, the younger 18 Sco varies by a few
percent more than the Sun, while the older stars 16 Cyg A and B vary a
few percent less.  The flat activity star $\tau$ Cet, also of significantly lower mass, has the smallest
range of variability of the sample.  Indeed, this star might be in a
state analogous to the solar ``Maunder Minimum'', a period of
temporarily reduced spot activity during a longer phase of cyclic
variations \cite[e.g.][]{Judge+others2004b}. Alternatively, it could
be in a permanent state of magnetic quiescence brought about by a
reconfiguration of the large-scale field \citep{Metcalfe2016}, which  may occur in all middle-aged sun-like stars \citep{vanSaders2016}.  Interestingly,
with spectral type G8~V, $\tau$ Cet is almost a K star. Main sequence K stars are highly likely to be found exhibiting cycling behavior \citep{Egeland2017}.

\subsection{Differential comparison of five stars}

Pertinent emission features in the spectra of these stars are shown in
Figure~\ref{fig:all}.  The figure shows the surface flux density
${\cal F}_\lambda = 1.70\times10^{17}/(\phi^\prime)^2\ f_\lambda $,
where $f_\lambda$ is the flux density measured at Earth.  The Figure
includes lines formed mostly in the solar transition region:
\ion{Si}{3} 120.6 nm, \ion{N}{5} 124.2 nm,
  \ion{Si}{4} 139.3 and 140.2 nm, \ion{C}{4} 154.8 and 155.0 nm. \revise{The 
  \ion{O}{5}] at 121.8 nm is also present, but shown in Figure~\ref{fig:sun}}.
  Lines
  of neutral carbon near 165.7 nm are highlighted, along with the
  multiplet of \ion{C}{2} at 133.4 and 133.5 nm.  Other lines of
  \ion{H}{1} L$\alpha$ (121.5 nm) and \ion{O}{1} (130.2, 130.4, 130.6,
  135.6, 135.9 nm) are contaminated by emission from the Earth's
  geo-corona, or are rather weak (\ion{N}{1} 131.9 nm).  The panel
  labeled ``O~V]'' in Figure~\ref{fig:all} contains broad emission in
    the wing of \ion{H}{1} L$\alpha$ 121.5 nm, with significant
    contamination from the geocorona in the COS data shown.

%
%

\figall

We consider first the four G1 to G3 V stars as a group, \revise{excluding the Sun, for which no comparable flux spectrum exists}.   Remarkably, the flux
spectra (i.e. spectra other than those of the Sun itself) are {\em
  qualitatively and qualitatively very similar}.  Let us consider the
lines in terms of increasing temperature of formation.  The purely chromospheric \ion{C}{1}
165.7 nm flux densities are within 10 percent of one another. 
The 
 \ion{C}{2} multiplet ($2s^2 2p\  ^2{\rm P}^o_ - 2s2p^2\ 
^2{\rm D}$) 
consists of three transitions,
the $J=1/2 \rightarrow J'=3/2$ 
line at 133.4532 nm, and the 
$J=3/2 \rightarrow J'=3/2,5/2$ 
blended lines at 133.5663 and 133.5778 nm.  The 
blended 133.57 nm 
profiles are almost identical except in the very cores of the
lines.  These lines are well resolved spectrally by COS and (for $\alpha$ Cen~A, STIS), as can be seen by the narrow width ($\approx 0.01$ nm) of the optically thin 135.56 nm chromospheric 
line of \ion{O}{1}] shown in Figure \protect\ref{fig:sun}.

While the \ion{C}{2} 133.45 nm line profiles are affected differently by interstellar
absorption,  the core flux densities of the 133.56 nm components all lie in the range
$1.3\pm0.2\times10^5$ \flxu{}~nm$^{-1}$. All the G2~V stars show small
self-reversals in the very core.  The profiles of 
lines of third, fourth and fifth spectra (first spectrum = neutral, second = singly ionized, etc.) all lie within 15\% 
of another.  For most lines the $\alpha$ Cen STIS spectrum is brightest, followed by
18 Sco, 16 Cyg A and 16 Cyg B.  Even the 164.0 nm (Balmer-$\alpha$)
transition of \ion{He}{2}, notoriously poorly understood, but probably
influenced by higher energy photons and/or particles from the corona
\citep{Laming+Feldman1993,Wahlstrom+Carlsson1994}, varies by less than
$15$\% of the average flux density.   \revise{The absence of detectable soft X-rays above 0.3 keV found below 
(section~\ref{subsec:younger}) does not invalidate the potential role of EUV photons in exciting helium, because helium 
photoionizations occur for $\approx 10\times$ softer photons at 0.02 and 0.04 keV for neutral and ionized helium respectively}

The 164 and 165 nm regions shown
in the figure include some UV continuum which are also remarkably similar
in brightness.  The UV continuum below 150 nm was shown to be a
tracer of magnetic activity and rotation for much younger stars by
\citealt{Linsky+others2012}.  
\revise{For each of 
 the collisionally-excited transition-region lines in Figure~\ref{fig:all} (i.e. except \ion{He}{2} and \ion{C}{1}), we divided the each flux by the average of the four stellar spectra, and computed the standard deviation of these normalized wavelength-integrated fluxes.  The standard deviation is just $\sigma=$10\%.}

These results stand in contrast to and strengthen (by extending the
sample to older and younger stellar spectra) the differential study of
the Sun, $\alpha$ Cen A (G2 V) and $\tau$ Cet (G8 V) by
\citet{Judge+others2004b}. In that work,  significant differences were found 
between the Sun and $\alpha$ Cen A on the one hand, and the older star
$\tau$ Cet.  These differences are evident as $\tau$ Cet is seen as the
low out-lier in brightness in the transition region lines in Figure
\ref{fig:sun}, and its lines of \ion{Si}{4} and \ion{C}{4} are
systematically blue-shifted relative to the other stars.

\subsection{Relationship to younger stars}
\label{subsec:younger}

To place these results in a broader context, we searched the list of
``solar analog'' stars (Table~4.1 of \citealt{Egeland2017}) for those
with useful UV data.  The following stars have data from HST or the
IUE satellite: HD 20630 ($P_{rot}$ =9.2 days, $L_{BOL}/L_\odot=0.82 $), 
HD 30495 (11.3, 0.96),
HD 72905 (4.89, 0.97),
HD 97334 (8.25,1.06), and HD 76151 (15.0,,0.97).
\revise{Fundamental data are based upon HIPPARCOS parallaxes and the work of  
\cite{Valenti+Fischer2005}, except for HD72905, where we used \cite{Gaidos+others2000}.}

\figyoung

Figure~\ref{fig:young} shows scaled flux densities 
versus age for the old stars together with younger
"solar analog'' stars with available UV measurements.  For each star,
a mean value of $R^\prime_{HK} = {\cal F}_{HK}/ {\cal F}_{BOL}$ is plotted (upper
panel), and the equivalent value $R^\prime_{UV} = {\cal F}_{UV}/ {\cal F}_{BOL}$ is
shown in the lower panels for the multiplets of ions of C, Si and He.
Here, ${\cal F}_{BOL} = L_\ast / 4\pi r_\ast^2$ is the stellar bolometric
flux density (\fluxu).  ${\cal F}_{UV}$ (also \fluxu) is the surface flux
density of the line or multiplet, minus the background continuum,
integrated over wavelength.  Notice the small range in chromospheric
$H$ and $K$ emission compared with the higher energy UV emission
lines.  The figure shows a systematic decrease of \ion{Ca}{2} and UV
$R^\prime$ values with estimated age.  While the $R^\prime_{HK}$ seems
to follow a smooth decline with age, the $R^\prime_{UV}$ values could
also be consistent with a stepwise drop in activity between 2 Gyr and
4 Gyr.  It must be noted that the \ion{Ca}{2} values for
$R^\prime_{HK}$ are subject to considerable empirical corrections for
low-activity stars, because the core emission is weak relative to the
bright underlying photospheric component. As is well known, this
increase in contrast of UV and X-ray data offer a
much cleaner measure of magnetic activity than the \ion{Ca}{2} data, at
the expense of having to observe from space.
Unfortunately, the competition for and relative rarity of satellite UV observations precludes the general use of UV data as proxies of magnetic activity for most stars.

The lowest panel of Fig. \ref{fig:young} shows soft X-ray surface flux densities, again scaled to each target's bolometric luminosity. The dashed line shows the ``minimum'' flux density of 10$^4$ erg cm$^{-2}$ s$^{-1}$, found by \citet{Schmitt1997} in the soft ROSAT bandpass, for a star with solar luminosity. Available measures of X-ray activity were obtained as follows: HD 72905 \citep{Ayres2017}; HD 97334 \citep{Linsky+others2013}; HD 20630 \citep{Ayres2017}; HD 30495 \citep{Wright+others2011}; HD 76151 \citep{Vidotto+others2014}; 18 Sco \citep{DeWarf+others2012}; a Cen A \citep{Ayres2017}; tau Ceti \citep{Ayres2017}; and the Sun \citep{Ayres2017}. 

For 16 Cyg A and B, X-ray activity measures have been acquired by us. In addition to the HST observations of 16 Cyg A and B, contemporaneous X-ray observations were carried out with Chandra on 7 Feb, 2016. 
Up to that time only upper limits were known. X-ray observations were obtained using the ACIS-I instrument for a total exposure time of 73.9 ksec. 
The data were reprocessed using CIAO v4.9 and reduced in the usual fashion. X-ray observations were secured for both 16 Cyg A and B, along with 16 Cyg C, the close ($\approx$3.6'' separation) M4.5V companion. 
\revise{Despite the long 73.9 ksec exposure time, the total counts for all targets were low.   In the ACIS-I energy range of 0.3-2.5 keV, the total counts for 16 Cyg A, B, C are just 43, 3, and 6  respectively, the  background noise is $\approx 1$ count.  In order to estimate thermal  coronal properties, 
single temperature MEKAL plasma models \citep{Mewe+others1995, Drake+others1996} were fitted to the energy distributions using the Sherpa modeling and fitting package, the results are listed in 
Table~\ref{tab:xray}. The X-ray fluxes of 16 Cyg B  more than an order of magnitude below 16 Cyg A. The results for B and C are less secure than for A. For component B it should be considered an upper limit.  This upper limit to $L_X$ lies among the lowest measured so far, for a solar-type star. The modeled temperatures in Table~\ref{tab:xray} are clearly open to further interpretation. It would help to re-observe this system.}

\revise{The \ion{Ca}{2} time series shown in Figure~\ref{fig:Sts} suggest that the X-ray data were acquired during a season with typical level of chromospheric activity}  seen on the A and B components 
 since 2008. \revise{ Also, lines such as \ion{N}{5}, generally believed to originate from plasma significantly heated by conduction from the corona, have very similar flux densities in A and B.  
The remarkably dim X-ray emission of the B component} is reminiscent of the behavior of the corona of $\alpha$~Cen~A, found via spectroscopy to be due to a modest cooling of the corona of this star \citep{Ayres+others2008}.   \revise{Existing soft X-ray missions are not quite `soft enough' to detect a corona which can still be seen in EUV emission (energies 
$\lta 0.2$ keV), routinely seen for decades 
in the solar corona in lines such as of \ion{Fe}{9} near 17 nm  ($\equiv 0.06$ keV).}

\begin{deluxetable}{lrrrr}
\tablecaption{X-ray Parameters of the 16 Cyg ABC System \label{tab:xray} }
\tablehead{ 
\colhead{Star}   &       \colhead{$kT$}    &     \colhead{$f_{\rm X}$}    &   \colhead{$L_{\rm X}$}     &    \colhead{$\log~L_{\rm X}$} \\
\colhead{}    &    \colhead{(keV)$^a$}    &   \colhead{(erg s$^{-1}$ cm$^{-2}$)}    &   \colhead{(erg s$^{-1}$)}  &  \colhead{(erg s$^{-1}$)}    }
\startdata
16 Cyg A         &            0.37  &  $1.1\times10^{-14}$  &  $5.7\times10^{26}$  &  26.7   \\
16 Cyg B    &  \ldots	&  $\lta 6\times10^{-16}$  &  $\lta 3\times10^{25}$  &  $\lta25.5$   \\
16 Cyg C                 &  $\approx$0.25  &  $\approx 2.6\times10^{-15}$  &  $\approx1.4\times10^{26}$  &  $\approx$26.1   \\
\enddata
In deriving these quantities, We use an interstellar column density 
$N_{\rm H} \approx 4\times10^{18}$ cm$^{-2}$ 
from \citet{Paresce1984}, with the Hipparcos distance of 21.1 pc.
\tablenotetext{a}{1 keV $\equiv 1.2 \times10^7$ K.}\end{deluxetable}

This stellar sample includes only stars with fundamental characteristics demonstrably
similar to the Sun \citep{Egeland2017}. Figure~\ref{fig:nl} shows ${\cal F}_{UV}$ against
${\cal F}_{HK}$ for the sample, roughly following  well-known non-linear
relationships found for very diverse samples of stars
\citep{Oranje1986,Oranje+Zwaan1985,Schrijver+others1985,Schrijver1995}. \revise{But
these linear fits in the log-log plane} are not justified any more than,
for example, a step function between the two groups of stars.  It is
adopted below merely to re-examine the power laws analyzed in the
above cited work.

\fignl

These data point to one conclusion: \emph{the level of non-radiative heating,
\revise{reflected in transition-region lines, converges to a level that is constant to within $\pm $10\%}, for G2 main-sequence stars older
than $\approx 2-4$ Gyr}.  

\subsection{The Sun as a star and its center-to-limb behavior}

No high resolution ($R \geq 10^4$) solar irradiance spectra exist, so
we must compare stellar data with solar data from individual features
on the Sun. Figure~\ref{fig:sun} shows spectral lines from the G130M 
flux spectrum of 18 Sco (there is to date no G160M
spectrum) and $\pi I_\lambda(\mu=1)$ where $I_\lambda(\mu=1)$ is the
quiet Sun intensity spectrum at disk center, at wavelength $\lambda$,
which we adopt from the atlas of \cite{Curdt+others2001}.  Here $\mu$
is the usual cosine of the angle between the local vertical on the Sun
and the line-of-sight to the observer.  The assumption of no limb
brightening/darkening -- a Lambert radiator -- is certainly incorrect
\citep{Dammasch+others1999, Worden+Woods+Bowman2001}. It is made here
to enable us to judge the departures from this approximation, and
thereby understand results in the literature that appear to confuse
flux-flux with intensity-intensity relations, which cannot be the same
(see below).  Assuming spherical symmetry, the flux density
$f_\lambda$ observed a distance $d$ from a star is:
\begin{eqnarray} \label{eqn:flux}
f_\lambda &=& \frac {R_*^2} {d^2}\ 2\pi \int_0^1 \! I_\lambda(\mu)\mu
d\mu,
  \label{eqn:i2flux}
\end{eqnarray}
where $I_\lambda(\mu)$ is the outward directed intensity just above
the stellar surface.   The surface flux densities used above are simply 
${\cal F}_\lambda = f_\lambda \frac {d^2} {R_*^2} = f_\lambda \  1.70\times10^{17}/\phi^{\prime 2} $.
``Flat'' limb behavior means $I_\lambda(\mu)$ is
independent of $\mu$, $\bar{I_{\lambda}}$, so that
\begin{eqnarray} \label{eqn:flux1}
f_\lambda &=& \frac {R_*^2} {d^2} \ \pi {\bar
  I_{\lambda}} \label{eqn:ibar}
\end{eqnarray}
Following \citet{Judge+others2004b} we write departures from this
``flat'' limb behavior in terms of $\beta_\lambda$:
\begin{eqnarray} \label{eqn:flux2}
f_\lambda &=& \frac {\pi R_*^2} {d^2}\ I_\lambda(\mu=1)
\ \beta_\lambda, \label{eqn:betanu}
\end{eqnarray}
\noindent where $\beta_\lambda$ is given by
\begin{equation}
I_\lambda(\mu=1) \ \beta_\lambda \ = \ 2\int_0^1 \! {I_\lambda(\mu)}
\mu d\mu.
\end{equation}
\noindent
Each individual wavelength $\lambda$ has its own center-to-limb curve
leading to its own $\beta_\lambda$.  But it is possible to define a
center-limb curve for an entire transition labeled $i$, in the case of
optically thin lines and for narrow ranges of continuum wavelengths.
\citet{Judge+others2004b} list quantities $\beta_i$ for the UV
continuum below the edge of \ion{Si}{1} at 152 nm, and
wavelength-integrated emission lines, from various sources.
\figsun
Typically, $\beta_i$ varies from close to $1$ for continuum and
optically thick transitions formed in the chromosphere, to $2.1$ for
thinner transitions formed in the less opaque transition region lines.
The large differences between 18 Sco and the solar spectrum $\pi
I_\lambda(\mu=1)$ evident in Figure~\ref{fig:sun} are at least partly
attributable to these center-to-limb effects\footnote{The data near
  121.8 nm in Figure~\ref{fig:sun} shows another emission line in the
  L$\alpha$ wing, which is one of the two resonance lines of
  \ion{Mg}{10} at 60.9794 nm, seen in SUMER's second spectral order,
  absent in stellar spectra.}.  It is likely that remaining systematic
differences between the 18 Sco and estimated solar 
flux spectrum exist because SUMER spectrum contains absolutely no
active regions at all \citep[Figure 2 of][]{Curdt+others2001}, and spotless days are quite rare on the Sun. This, in spite of the fact that the COS spectrum of 18 Sco was observed in February of 2011, when the
$S$-index of the star was measured to be at a minimum on decadal time
scales, some 10\% lower than typical values.

Within statistical uncertainties, we conclude that {\em the UV flux
  spectrum of 18~Sco is a useful proxy for the
  Sun-as-a-star.}  It is a pity that no high quality Sun-as-a-star
spectrum in the commonly observed vacuum UV region has been obtained.

\subsection{De-convolution and detailed analysis of line profiles}

The COS instrument has a ``line spread function'' (LSF) with a
well-documented core-halo structure \citep{Ghavamian+others2009}.
Considering only the core component, the COS instrument's spectral
resolution is $R \approx 25,000$ ($\equiv 12 $ \velu{} in Doppler units).  
This resolution is more than
sufficient to explore other signatures of magnetic activity that are
present in the Sun and Sun-like stars.  \citet{Wood+Linsky+Ayres1997}
measured
broad components of transition region lines in 
a wide 
variety of late-type stars, finding widths in the dwarf stars 
approaching $10^2$ \velu, 
at the base of transition region lines.  The only Sun-like stars in their sample 
are $\alpha$ Cen A and the Sun itself.  In the Sun, these components result from
``explosive events'', discovered with the HRTS instrument
\citep{Brueckner+Bartoe1983, Dere+Bartoe+Brueckner1989, Dere1994}.
\citet{Wood+Linsky+Ayres1997} found that in spectra of $\alpha$ Cen A,
the broad components of lines of \ion{Si}{4} amounted to $\approx 1/4$ of the
total emission. In contrast, they estimated an upper limit of 1/20 for
the Sun. 
\citet{Wood+Linsky+Ayres1997} were therefore  hesitant to identify the $\alpha$
Cen profiles with explosive events. They termed the phenomenon 
microflares.  As noted above, these authors had no choice but to compare stellar flux data with solar intensity data. With no knowledge of center-to-limb behavior of both components, this result is open to question. 

Unfortunately, broad stellar components are qualitatively similar to
the wings present in the LSF of COS. But the LSF's are very well
known.  De-convolution is therefore possible.  Some additional
``smoothness'' constraint is needed to constrain the de-convolved
profiles, because they are non-unique in the presence of noise.
Assuming that the LSF's are precise, we have de-convolved the COS
spectra using the wavelength-dependent LSF's appropriate for each
wavelength region.  We used the procedure {\tt max\_entropy.pro}
released with the standard {\em SolarSoft} library to regularize the
inversions.  The presence of noise in the spectra (signal-to-noise ratios are 
$\lta 30$) enforces some subjectivity in the application of
such algorithms, as one must balance the regularization parameter
(in this case the maximum permitted entropy of the recovered data) with
the need to preserve spectral structure in the data.  We 
found that 6 iterations of the algorithm were sufficient to provide 
a balance between these factors.   The results are robust against
different deconvolution parameters, including a maximum likelihood scheme.

\figdeconv

The results are summarized in Figure~\ref{fig:deconv}.  The upper panels show
measured and deconvolved COS data as well as STIS data for $\alpha$ Cen A.  The second and third rows
show fits of double-Gaussian profiles and Voigt profiles to the deconvolved 
data.   
The data are shown for the \ion{Si}{4} line
at 139.3755 nm, results for other lines are
of lower quality owing to smaller S/N ratios, except for \ion{Si}{3} which is probably affected by photon scattering.

The values of ``chi$^2$'' listed in the figures,  \revise{reduced-$\chi^2$ values}, were derived from 
the noise reported by the HST software.   The results as shown 
are clear.   The STIS data for 
$\alpha$ Cen A shows clear evidence of Gaussian broad profiles 
as documented by \citet{Wood+Linsky+Ayres1997}. The other stars for which 
only COS data are available, do not.  $18$ Sco, 16 Cyg A and B show 
profiles more consistent (according to $\chi^2$ values) 
with Voigt than double-Gaussian profiles. Therefore we conclude that 
either:
(1) the COS deconvolutions are simply too unreliable to be trusted, or 
(2) 18 Sco and the 16 Cyg stars genuinely show no detectable evidence of
the ``microflare'' phenmenon reported by \citet{Wood+Linsky+Ayres1997}. 

Similar results (not shown) are found for the 
 \ion{Si}{3} 120.6 and \ion{Si}{4} 140.2 nm lines, in which the Voigt profiles 
are vastly superior to the double-Gaussian fits in every  COS spectrum, except
for $\alpha$ Cen~A, which reveals a broad component 
near 20 and 15\% of the peak fluxes respectively.  However, the derived values of
the Voigt parameter ($a = \gamma / 4\pi \Delta \nu_D$, $\gamma=$ damping width, 
$\Delta\nu_D$ =
Doppler width, both in frequency units) are much larger than 
can be accounted for by collisions or other physical effects in the emitting plasmas.
($\gamma$ is given by ``damp'' and $\Delta\nu_D$ by Dopp in Figure~\ref{fig:deconv}).
Perhaps 
residual instrumental effects are present even after the deconvolutions. 
The 120.6 nm line of \ion{Si}{3} is
significantly broader than the other stellar lines, suggesting some \revise{stellar}
opacity broadening in addition to Doppler motions, which may influence 
the interpretation.

\revise{On balance, it seems safe to conclude that HST spectra indicate
 the presence of microflare events in $\alpha$ Cen A, but any phenomena comparable on  
  18 Sco, 
16 Cyg A or B remain undetected.  As a conservative upper limit, 
we can assume the broad components must have a peak brightness $\leq 30\%$ of the narrow peak, simply by adopting the parameters of
the double-Gaussian fits which have far higher reduced-$\chi^2$ values than their Voigt counterparts. The difference between 
the $\alpha$ Cen~A spectrum and disk spectra of solar explosive events remains unresolved, and will probably do so until genuine Sun-as-a-star measurements become available. }

\subsection{Unreasonable agreement? }

The results highlighted in Figures \ref{fig:all} and \ref{fig:sun} are, 
in a particular sense, 
worrisome.  The measured flux densities of five Sun-like stars lie
within $10$\% of one another, and the line profiles
are remarkably similar.  This result, initially gratifying, appears on
deeper examination to be unsatisfactory, because of two related facts.
\begin{enumerate}
\item \revise{Well-known statistical correlations exist
  between fluxes of chromospheric (${\cal F}_{ch}$), and transition region
  and/or coronal lines (${\cal F}_{z}$) \citep{Ayres+Marstad+Linsky1981,
    Oranje1986}. These take the form}
\begin{equation} \label{eq:gamma}
{\cal F}_{z} = {\cal F}_{ch}^\gamma,\ \ \gamma \ge 1.
\end{equation} 
These relations come from scatter plots of (mostly) single
measurements of individual stars.
\item At the same time we also know that fluxes of chromospheric
  multiplets such as the \ion{Ca}{2} $H$ and $K$ lines, vary
  significantly as a function of rotational and sunspot cycle phases
  (see Figure \ref{fig:Sindex}).  For the Sun itself, the smoothed
  \CaII{} $S$-index varies from 0.162 at minimum to 0.177 (a change of +9\%), with
  much larger excursions when including variability on shorter
  time-scales ($\sim$0.156-0.190, 22\%; \citealt{Egeland+others2017}).
\end{enumerate}
The problem is that if we differentiate equation~(\ref{eq:gamma}), we
find
\begin{equation} \label{eq:differ}
\delta \ln {\cal F}_{z} = \gamma \ \delta \ln {\cal F}_{ch}.
\end{equation}
Assuming  that sampling an ensemble of stars is
equivalent to seeing variations of a typical star (i.e., that
equations~\ref{eq:gamma} and \ref{eq:differ} apply to each individual star), then {\em
  small fractional changes} in chromospheric emission ($\delta \ln {\cal F}_{ch}$)
should lead to {\em much larger fractional changes} in the
corresponding fluxes in transition region and coronal lines ($\delta \ln
{\cal F}_{z}$). But the measured flux UV densities ${\cal F}_{UV}$ of the 
five low-activity 
Sun-like stars lie
within $\pm 10$\% ($\pm1\sigma$) of one another.  

\revise{Let us} assume (from the definition of $S$) that $S$-indices for such a homogeneous group of stars are linearly
related to ${\cal F}_{ch}$, and for ${\cal F}_z$ we use the \ion{Si}{4} 139.3 nm
flux, because it has high S/N ratios.  For this pair of lines the
value of $\gamma$ can be estimated from known logarithmic dependences of
\ion{Mg}{2} (chromospheric) and transition region lines, where
\citep{Oranje1986}, $\gamma \approx 1.3$.  Using data from the F8 and
G type dwarf stars in Table 2 of \cite{Oranje+Zwaan1985} we would find that
${\cal F}_{Mg~II} \propto S_{Ca~II}^{1.8}$, where $S$ is the
``$S$-index''. Thus we expect that 
\begin{equation} 
\label{eq:dfi} {\delta
    \ln {\cal F}_{Si~IV}} \approx 
    1.3\times1.8 \ {\delta \ln S_{Ca~II}} = 
    2.4 \ {\delta \ln S_{Ca~II}}.
\end{equation}
Here we must only consider the non-linear relationship given by
equation~(\ref{eq:gamma}), without ad-hoc corrections for ``basal''
fluxes \citep[e,g,][]{Schrijver1995}. \revise{Note that equation~(\ref{eq:dfi}), based upon previously 
published work, also applies 
roughly to our smaller sample with different ages, where $\gamma \approx 2.35$ (Figure~\ref{fig:nl}). }

\revise{The LHS of 
 equation~(\ref{eq:dfi}) amounts to 10\%, or 0.1 in $\ln \cal{F}_z$.   To evaluate the RHS, we must estimate 
 variations in $S$-index within each season's observation that coincides with the epoch of UV observations. These amount to 2-4\%.  But, based upon high cadence solar observations, these values must be treated as strict lower limits since the time-series there are large gaps. 
Then, with $\delta \ln
S \gta 0.04$, we  expect }from
(equation~\ref{eq:differ}) that 
\begin{equation} \label{eq:gammasi}
{\delta \ln {\cal F}_{Si~IV}} \gta 0.1
\end{equation}
for each star.  In pairwise comparisons of fluxes between stars, we must multiply by $\sqrt{2}$.
Therefore we should observe
  (random) differences between the measured flux densities \revise{with an expectation variation
  of {\em at least} $14\%$. 
It is in this sense that {\em this agreement is perhaps 
  unreasonable}. } 
 
\revise{Given the small numbers of stars and the single-epoch stellar UV observations, the disagreement cannot be considered statistically significant.  However,  
in the Appendix we show how essential physics of the passage of
active regions across  stellar disks influence the interpretation of such data in the future.  The new 
elements include
the different geometry of formation of chromospheric and transition-region (and coronal) lines, and 
the very different center-limb variations of these features.} 
These considerations have significant implications 
for ``flux-flux'' relations and ``basal fluxes'' \citep{Schrijver1995} as
discussed below.

\section{Discussion}

We have compared UV emission line spectra of Sun-like stars from
age 0.2 to 7.6 Gyr.  The first significant result is shown in Figure~\ref{fig:Sindex}, namely that for stars of solar mass (excluding $\tau$ Ceti) the median S indices seem to drop with age.  Equally signficant though is the observation that the variances in S index fall with age, in the latter half of the main sequence.  

In contrast, our UV observations, mostly from one obseration of each star, show that 
the bright cores of emission line profiles are
remarkably similar for all the $> 4$ Gyr G2-G3 stars studied here, independent of age, and
measurably different from $\tau$ Cet
\citep{Judge+others2004b}.  We have suggested an explanation for why such similar core UV profiles exist, given the statistical power-law relationships
between UV and \ion{Ca}{2} flux densities, 
and random sampling from measured  variations of \ion{Ca}{2}.  

The broad components of the
transition region lines examined in $\alpha$ Cen A by
\citet{Wood+Linsky+Ayres1997}, confirmed here, are not present in the COS spectra of 18 Sco and 16 Cyg A and B, as judged by the far smaller $\chi^2$ goodness-of-fit measures for Voigt, not double-Gaussian, profiles. 
Interestingly, 
of the five stars under study,
$\alpha$~Cen A and $\tau$ Ceti have the highest and lowest masses and
radii.  The data do not support the idea that 
 broad components 
decay significantly in time from 4 Gyr
to 7.6 Gyr, which would otherwise indicate 
that explosive events will become rarer as the
Sun ages on the main-sequence.
Although Voigt profiles provide quantitatively better fits than 
double Gaussian profiles that have been used previously
\citep{Wood+Linsky+Ayres1997,Peter2000}, the unphysically large 
Voigt parameters
$a \gta 0.1$ seem to point to residual  effects from the COS instrument, or (less likely)  weak blended emission lines, 
improperly accounted for by the deconvolution.  Higher quality data
would be worth obtaining, including a first G160M spectrum of 18 Sco.

The origin of the significant spectral differences between $\tau$ Ceti and the other stars 
cannot be
identified uniquely. $\tau$ Ceti  has a lower mass, older age, and it may be seen almost pole-on
\citep{Judge+others2004b}.  Previous work, based
upon weak variations in $S$-index 
\citep{Baliunas+others1995}, suggested that these spectra could reflect
conditions present during a ``Grand Minimum'' episode
\cite{Judge+others2004b}. But both 16 Cyg A and B have ``flat''
$S$-index activity records, with quite different spectra from those of $\tau$ Ceti. These findings cast
doubt on the proposal than flat activity in stars like $\tau$ Ceti is really
 associated with a solar-like ``grand minimum''.

\revise{The Sun typically exhibits \~10-15\% variations in chromospheric activity 
over its ~11 year activity cycle (Egeland et al. 2017). 
However, as shown by Ayres (2014, and references therein), the Sun's coronal X-ray emission (~0.2-2.5 keV) changes by ~7-8 times over its activity cycle. 
For 16 at least Cyg A (and perhaps B), the X-ray emissions indicate that magnetic dynamos are still operating in these older suns. The very low $L_X$ observed for 16 Cyg B might be explained if 16 Cyg B was observed near the  minimum state of an X-ray coronal cycle, or a large magnetically unipolar region (coronal hole) might dominate the visible hemisphere of 16 Cyg B during the Chandra visit.
We plan to request additional X-ray observations of the 16 Cyg system using the enhanced soft response of the Chandra HRC, to determine possible short (rotation driven) and long-term variations in coronal activity of 16 Cyg B and its companions.}

A corollary of our work (in the Appendix) is that, when considering
``flux-flux'' relationships \citep{Oranje1986,Oranje+Zwaan1985}, one
cannot mix spatially-resolved data for the {\em intensity} of spectral
features with those of unresolved {\em fluxes} from stellar disks, as
done by \citep{Schrijver+others1985,Schrijver1992,Schrijver1995} for
example.  This is most easily seen by considering the passage of one active
region across the disk of the Sun or a star.  The \ion{Ca}{2} $S$-index,
sampling a mix of chromospheric (core) and photospheric (wing)
emission, forms under optically thick conditions in which limb
darkening occurs (radiation is peaked in the radial direction).  Lines
such as \ion{Si}{4} form under much thinner conditions and are limb
brightened.  This means that the same active region will exhibit one
ratio of $I_{Si~IV}/I_{Ca~II}$ near the limb and another at disk
center.  The solar active region McMath 12488 analyzed by
\citet{Schrijver+others1985,Schrijver1992} sampled disk-center data
only, with $0.92<\mu\leq1$.  These data were used by Schrijver to draw intensity-intensity
 relationships on the same footing as flux-flux relationships.  But the analysis is on shakey grounds, for 
if these analyses had been performed
close to the solar limb, entirely different results would have been
found, invalidating the entire 
approach (see the Appendix). 
The different center-to-limb- 
intensity variations translate into time-dependent rotational
modulations in disk-integrated fluxes which are quite different, depending only on the
(observer-dependent) $\mu-$angle of the active region.

The question arises as to the evolution of the broad
features, perhaps analogous to solar explosive events, from 4 to 7 Gyr.  There is no clear one-to-one  correspondence with age in our data, since 18 Sco and 16 Cyg A and B (no detectable broad regions) and $\alpha$ Cen A 
(20-25\% broad region contribution) 
would indicate an increase and then 
decrease of broad emission features with age. 
As documented by
\citet{Dere+Bartoe+Brueckner1989}, solar explosive events seem to
avoid active regions and peak at mid latitudes in association with the
bright chromospheric network where magnetic field is concentrated in
quiet regions by supergranular advection.  They are equally present in
coronal holes as in quiet regions, indicating that the large-scale
pattern of magnetic field is unimportant for the generation of such
events.  Later, \citet{Dere+others1991} ``identified [explosive
  events] with magnetic reconnection that occurs during the
cancellation of photospheric magnetic flux,'' associated with small-scale emerging
flux during the formation of active regions, as well as along
supergranulation cell boundaries.

More data of higher quality appear necessary to probe 
the evolution of surface magnetism on main sequence Sun-like stars.  The broad-line regions are important because they directly reflect magnetic processes occurring significantly above the sound speed in the emitting plasma.

\section{Summary and outlook}

UV spectra of the main-sequence solar mass stars reflect the changing surface magnetic fields as a function of age.  Even for stars of 7 Gyr age, the data reveal lines of
four-times-ionized \ion{O}{5}] and \ion{N}{5}, which, outside of active region loops are powered via conduction
  of heat downwards from an overlying corona  \citep{Jordan1980b,Jordan1992}.  The emerging 
picture is that while emerging magnetic flux and energy from beneath
the surface decays rapidly with time up to between 1 and 2 Gyr  \citep{Skumanich1972,Guedel2007, vanSaders2016}, the
magnetic energy re-radiated 
at UV wavelengths converges to a specific
fraction of the star's luminosity to within  $\approx 10\%$ after this age.
Given the known variations of several tens of percent in UV line radiation
with the solar cycle, this result was initially surprising, 18 Sco certainly supports a healthy magnetic activity cycle, yet the UV spectra are very similar to the non-cycling 16 Cygni stars.  It is notable that
the $S$-indices of all stars at around the epochs of the UV observations
are close to 0.16 to 0.17, suggesting that we have (by chance)
obtained data close to minimum levels of activity of at least 18 Sco
and $\alpha$ Cen A.  If this is the case, then our observations
reflect the operation of a residual local and small scale,
``turbulent'' dynamo, in the presence of convective motions in a
highly non-diffusive plasma (high magnetic Reynolds number), even in the
absence of a large-scale dynamo driven in part by global stellar
differential rotation \cite[e.g.][]{Voegler+Schuessler2007,Lites2011}. 
%
%
The presence of a large-scale variable magnetic field associated with
decadal sunspot variations on the Sun is rare among G-type stars. Such
cycling behavior is far more common among K stars on the main-sequence
\citep{Egeland2017}.  Neither 16 Cyg A or B, with ages of 7 Gyr,
exhibits cycling behavior in the $S$-index, yet their rotation periods
are slightly smaller than for the Sun, and Rossby numbers very similar.  
This begs the questions: What
causes cycling behavior in the first place? Will the future Sun
exhibit current cycling behavior as it ages on the main-sequence?

Lastly, it seems important to obtain genuine 
spectra of the Sun-as-a-star with spectral resolutions in excess of $10^4$, in order to make use of the rich UV archives of high quality UV data obtained with the Hubble Space Telescope.  Such work is currently being proposed (C. Kankelborg, P. Judge and colleagues).

\acknowledgments Acknowledgments.

We are grateful to Manuel G\"udel for very helpful comments
on the manuscript.  
\facilities{Hubble Space Telescope, Solar and Heliospheric
  Observatory, Mt. Wilson $H$ and $K$ survey data, Lowell Observatory
  Solar Stellar Spectrograph, SIMBAD, IUE observatory.}  \software{IDL Solarsoft}

\bibliographystyle{yahapj} \bibliography{references}

\appendix

\section*{Center-to-limb effects on rotational, cyclic modulation, and flux-flux relationships}

The ``disagreement'' found in the text is resolved through
consideration of center-limb variations as active regions cross the
stellar disks.  Let us consider \ion{Ca}{2} for which $\beta_i \approx
1$, and \ion{Si}{4} which has $\beta_i \approx 2.2$, and the passage of
one active region across the stellar disk. When $\beta_i \leqslant 1$,
the spectral feature is flat or limb-darkened; when $\beta_i > 1$, it
is limb-brightened, a value of $\beta_i=2$ implies that $I_i(\mu)
\propto \mu^{-1}$. This behavior also corresponds to an optically thin
plane-parallel atmosphere.

\figcl

As well as the background quiet Sun, We need to estimate the
center-to-limb behavior of typical active regions.  Unfortunately,
active regions evolve significantly on time scales of a half-rotation,
nor can we use existing (simultaneously measured) data from different
vantage points in the solar system to determine center-limb behavior,
because no imagers have obtained data for typical transition region
lines\footnote{The STEREO mission observes \ion{He}{2} 30.4 nm
  emission.  The problem is that helium lines are atypical \citep[see,
    e.g.][]{Pietarila+Judge2004}.}.  Therefore we must use theoretical
ideas to determine the C-L behavior of an active region.

\figpassage

The excess emission from chromospheric lines tends to form under
optically thick conditions in plasma that lies within stratified
layers of the atmosphere \citep{Linsky+Avrett1970}.  This is
especially true at the wavelengths contributing to the $S$-index of
\ion{Ca}{2} $H$ and $K$, where at least some of the emission passing
through the $S$-index filter profile forms in the upper photosphere, in
regions where there is a significant Wilson depression. The latter is
important because radiation emerges from magnetic regions depressed
below the average height.  These conditions lead to $\beta \leq
1$. This means that the radiation is mostly {\em radially directed},
and the flux contributed by the AR is reduced by the fore-shortening
of the projected area when close to the limb.  As an active region
crosses disk center, the contrast of the active region with its
surroundings reaches a maximum.

In contrast, transition region and coronal lines form under far less
thick conditions.  In active regions, contributions to the coronal
emission come from loops or from emission at the base of hotter loop
structures.  The same is true for transition region emission.  The sum
of all loop emission in an active region tends not to be
fore-shortened and can be represented by the assumption of
isotropic intensity. In contrast the foot-point emission can be fore-shortened
(area $\propto \mu$) and limb-brightened according to $I(\mu) \propto
\mu^{-1}$ if almost a plane-parallel emission region.  In both cases
the product of $I(\mu)$ and the area contributing to flux is the same.
Thus, in the limit of optically thin emission, the emission is
isotropic, which leads to no center-to-limb variation (the emitting
plasma overlying the active region emits the same power into all
directions independent of heliocentric cosine $\mu$).  This is
radically different from the thick (chromospheric) case, owing to both
fore-shortening and limb darkening.

To illustrate these differences we show in Figure~\ref{fig:passage}
the theoretical passage of an active region for one complete solar rotation, as
it affects the total flux density ${\cal F}$ for the thick and thin
cases. The active region is assumed to be a disk (chromosphere) or
sphere (for convenience, since for optically thin emission the
geometry does not matter except very close to the limb), both covering 5\% of the stellar surface.  Phase zero is chosen to be when the active region is on the central meridian. 
The chromospheric intensity contrast (active/quiet intensity at disk
center) is set to a factor of two, the corona an enormous (!) factor
of 100.  (The exact numbers do not matter so long as the intensities
follow non-linear relations of the kind measured for the Sun,
\citealp[e.g.][]{Schrijver1992}).  

The essential point of the figure is that the
\emph{modulations of integrated flux} (about 10\% 
for \ion{Ca}{2}, 68\% for the ``coronal'' case) are \emph{vastly different from the modulation 
of intensity}.  When spatially resolved, the active region is only twice as bright in \ion{Ca}{2} but is 100$\times$ as bright in the ``coronal'' case!   
This figure therefore explains not only the ``unreasonable agreement'' in the main text,
but has wider implications.  The passages of active regions across
disks of Sun-like stars are almost as easy to detect in limb-darkened
transitions such as \ion{Ca}{2} lines against a limb-darkened disk, as
they are in thinner transitions in the corona and transition
region. The latter show a far smaller flux contrast than the intensity
contrast of the active regions.  Along with previously noted problems
concerning the energy sensitivities of various instruments
\citep{Ayres1997,Judge+Solomon+Ayres2003}, the calculations also
explain the puzzle of why stellar activity cycles seen in fluxes of
X-rays are of modest amplitude given the amplitudes in
$S$-indices. This, in spite of the fact that solar active regions are so
much intense in X-rays than quiet Sun.

Finally, the figure shows that 
more than 50\% of the rotational phase space is
filled with high flux contrast as optically thin emissions (``corona'' in the figure) cross the
disk.  In comparison, the limb-darkened, fore-shortened thick flux
contrast peaks with width in rotational phase a factor $\geq$ two or more smaller. This means
that in the presence of several active regions on the stellar disk,
distributed in longitude, there is a higher probability that the
``thin'' cases will always have contributions from the active regions,
potentially washing out the observability of flux modulation due to rotation compared with the ``thick'' cases.

\end{document}